\documentclass[12pt]{article}
\usepackage[margin=1in]{geometry}
\usepackage[utf8]{inputenc}
\usepackage{natbib}
\usepackage{makecell}
\usepackage{pbox}
\usepackage{soul}
\usepackage{hyperref}
\usepackage{lipsum}
\usepackage{longtable}
\usepackage{amsmath,amssymb}
\usepackage{etoolbox}
\usepackage{float}
\usepackage{graphics, color}
\usepackage{bm}
\usepackage{graphicx}
\usepackage{epsfig}
\usepackage{makeidx}
\usepackage{comment}
\usepackage{float}
\usepackage{mathtools}
\usepackage{kotex}
\usepackage{siunitx}
\usepackage{booktabs}
\usepackage{multirow}
\usepackage{array}
\usepackage{tabularx}
\usepackage{lscape}
\usepackage{rotating}
\usepackage[labelfont={sf,bf}]{caption}
\usepackage{subcaption}
\usepackage{ulem}

\usepackage{enumerate}
\makeatletter
\patchcmd{\@makecaption}
  {\parbox}
  {\advance\@tempdima-\fontdimen2} % decrease the width!
  {}{}
\makeatother
\allowdisplaybreaks

\newtheorem{assumption}{Assumption}
\newtheorem{theorem}{Theorem}
\newtheorem{example}{Example}

\newtheorem{lemma}{Lemma}
\begin{document}
\def\spacingset#1{\renewcommand{\baselinestretch}%
{#1}\small\normalsize} \spacingset{1}

  \title{\bf \large Nearest neighbor ratio imputation with incomplete multi-nomial outcome in survey sampling}
  \author{\small Chenyin Gao   \thanks{ Department of Statistics, North Carolina State University, North Carolina
27695, U.S.A. Email: cgao6@ncsu.edu},\hspace{.2cm}Katherine Jenny Thompson\thanks{U.S. Census Bureau, 4600 Silver Hill Rd, Washington, DC 20233, U.S.A. E-mail: Katherine.J.Thompson@census.gov},\hspace{.2cm}Shu Yang \thanks{
   Department of Statistics, North Carolina State University, North Carolina
27695, U.S.A. Email: syang24@ncsu.edu},\hspace{.2cm}and Jae Kwang Kim\thanks{
  Department of Statistics, Iowa State University, Ames, Iowa 50011, U.S.A. Email: jkim@iastate.edu.}}
  \date{}
\maketitle

\bigskip
\begin{abstract}
    Nonresponse is a common problem in survey sampling. Appropriate treatment can be challenging, especially when dealing with detailed breakdowns of totals. Often, the nearest neighbor imputation method is used to handle such incomplete multinomial data. 
    In this article, we investigate the nearest neighbor ratio imputation estimator, in which auxiliary variables are used to identify the closest donor and 
    the vector of proportions from the donor
    is  applied to the total of the recipient to implement ratio imputation. 
    To estimate the asymptotic variance, we first treat the nearest neighbor ratio imputation as a special case of predictive matching imputation and apply the linearization method of \cite{yang2020asymptotic}.  
    To account for the non-negligible sampling fractions, 
    parametric and generalized additive models are employed to incorporate the smoothness of the imputation estimator, which results  in  a valid variance estimator. We apply the proposed method to estimate expenditures detail items based on empirical data from the 2018 collection of the Service Annual Survey, conducted by the United States Census Bureau. Our simulation results  demonstrate the validity of our proposed estimators and also confirm  that the derived variance estimators have good performance even when the sampling fraction is non-negligible. 
\end{abstract}
 \noindent%
{\it Keywords:}   Non-response, Hot deck imputation, Generalized Additive Models.
\vfill

\newpage
\spacingset{1.5} % DON'T change the spacing!
 
\section{Introduction} 
Sample surveys are often designed to estimate totals (e.g., revenue, earnings). However, in addition to collecting this information, many surveys request and produce sets of compositional variables (details) that sum to a total, such as a breakdown of total expenditures by type of expenditure or a breakdown of total income by source. Examples from the United States Census Bureau include the 2020 Current Population Survey’s Annual Social and Economic (ASEC) Supplement which defines total household earnings as the sum of total wages and salaries, farm income, and self-employment income and the 2018 Services Annual Survey (SAS) which requests detailed breakdown of the sampled business' reported total expenditures by expenditures on annual payroll, fringe benefits for employees, and expenditures on software, among other categories.  

This paper is concerned with the missing data treatment for these sets of detail items. In contrast to collected totals items, reliable auxiliary variable data for \emph{all} detail items are rarely available for all requested compositional variables.  Furthermore, there are often true zeros that differ across units. These limitations make it difficult to develop feasible parametric imputation models for each individual detail item and motivate the usage of hot deck imputation instead. Hot deck imputation -- sometimes called ``donor imputation'' -- obtains replacement values for nonresponding (or missing) data items by matching a donor record containing valid data to a recipient record containing invalid or missing data, imputing the missing values from the donor record \citep{andridge2010review,beaumont2009variance}.  {\citet{Kalton1982Kaspryzk} recommend using the same donor to impute sets of compositional details to preserve inter-item relationships in the multinomial data. To ensure additivity as well}, it is sensible to use some form of hot deck imputation to impute the \emph{vector} of proportions from a matched respondent (donor), then derive the imputed values for each imputed item by multiplying the (donated) proportion by the nonresponding recipient’s total \citep{bankier2000generic,little2008hot,andridge2015assessing}. 

If the proportions associated with each detail item within a imputation class appear to be approximately the same for each unit, then the donor can be selected at random within imputation cells. However, it is possible that the multinomial distribution of the details could be related to unit size. In this case, the unit size should be incorporated into the hot deck matching procedure. The simplest version selects the “nearest” donor using a suitable distance function such as unit size or -- as in this case -- the total, assumed available for all donors and recipients. Since the total is univariate, the absolute value of the difference is a natural distance function, yielding estimates that are asymptotically unbiased \citep{yang2020asymptotic}. Hereafter, we refer to the imputation process that selects the nearest neighbor as donor and imputes the sets of donor proportions as the nearest neighbor ratio imputation (NNRI) method.

Consider the Service Annual Survey (SAS), the subject of the empirical application presented in Section \ref{sec:application}. This program collects aggregate and detailed revenues and expenses from a stratified sample of business firms with paid employees in selected industries in the services sector. Given a lack of auxiliary data, weak historic reporting, and a high reported zero rate, developing good predictive models for each individual detail item collected by the SAS is perhaps infeasible. That said, there are verifiable predictors of the \emph{set} of detail items i.e., the multinomial distribution, specifically the industry in which the firm is classified, the tax-exempt status of the firm, and the size of the firm as measured by total revenue or total expenses. For example, a finance business will often report high proportions of total expenses in all three personnel cost categories, a scientific business is unlikely to report costs from temporary or leased employees, and a full-service restaurant would likely report most of its expenses from gross payroll and expensed equipment, materials, costs, or supplies. Consequently, imputation classes in the SAS are defined by industry code and tax-exempt status. As regards unit size, larger businesses are more likely to expense costs and track depreciation than a smaller business in the same industry. Accordingly, one would expect zero or nearly zero proportions of costs and depreciation values from smaller businesses, and positively valued proportions for the same component items from the larger businesses.  In turn, the proportion of total expenses represented by gross annual payroll tends to decrease as unit sizes increase.

In practice, it is often extremely difficult -- if not impossible -- to delineate exactly where the changes in multinomial distributions occur. However, using nearest neighbor donor selection procedure implicitly accounts for these subtle shifts. 

The primary purpose of imputation is to ``fill in the blanks with plausible (i.e., realistic) and consistent values" \citep{Sande1982}. From a bias reduction perspective, there are advantages in preserving reported and previously imputed totals as is done with NNRI. Auxiliary data are often available for direct substitution of missing or invalid totals \citep{Beaumont2011ONVE} or for modeling. Consequently, reported totals are generally validated (edited) and imputed when necessary early in the editing process, and missing values are imputed with very strong models. Not only are they made available for all units for subsequent hot deck imputation, such totals are usually ``goldplated" against further changes \citep{sigman1997}.

Fundamentally, totals are considered to be more reliably reported than sets of compositional details in the survey methodology literature, especially when the totals have accounting or financial record-keeping definitions (e.g., income, total sales, total expenses).  In contrast, the queried sets of compositional details are generally collected for \emph{statistical reporting purposes}, not accounting purposes, and are therefore not readily available \citep{Snijkers2013}. Indeed, it is likely that the reported percentage distributions are more reliably reported than the values themselves. For example, when interviewing a non-probability sample of large businesses, \citet{willimack2010nichols} learned that ``company reporters resorted (sic) to estimation strategies rather than leaving items unreported (i.e., blank). Moreover, they only used estimation schemes when company data did not include the type of detail requested on the report." This dovetails with the findings presented in \citet{andridge2015assessing} via a proxy pattern-mixture model analysis of selected items collected by the SAS: when annual payroll was used as the single predictor of total sales in a ratio imputation model, the fraction of missing information (FMI) values were close to zero, indicative of extremely accurate imputation models, whereas the FMI values for collected detail items using total sales as the sole predictor were near one (the maximum value). Finally, deriving hot deck imputed values by multiplying the recipient's total by the corresponding nearest-neighbor ratio is frequently used in business surveys \citep{beaumont2009variance}. In such positively skewed distributions, donating a ratio instead of a total guards against substitution of an overly large or small value from the respondent "nearest neighbor." Of course, this phenomena is not confined to business surveys. For example, the 2019 ASEC reports the U.S. median income as $\$68,703 \pm (\$904)$; the $95^{th}$ percentile is $\$270,002 \pm (\$4,831)$. See \url{https://www.census.gov/content/dam/Census/library/publications/2020/demo/p60-270.pdf}. 

Estimation from nearest neighbor ratio imputed data is straightforward. Variance estimation is less so, in part because the donor selection procedure is deterministic. The variance estimation is further complicated in our setting due to the potential shift in multinomial distributions as described above. \citet*{andridge2020finding} investigates multiple imputations of proportions with nearest neighbor ratio hot deck imputation using the Approximate Bayesian Bootstrap, finding consistent underestimation of the variance. As in the cited reference, we assume that the \emph{set} of details is reported or is missing; in practice, detail components that do not sum to their associated total and are not within a small raking tolerance are often treated as missing, and \emph{all} detail items are imputed, regardless of their original reporting status.  

Statistical inference under nearest neighbor imputation in survey sampling has been discussed by \cite{chen01}, \cite{shao2008}, \cite{kim2011}, \cite{yang2019nearest}, among others.  
In this paper, we discuss statistical inference under NNRI in survey sampling. To make statistical inference,  we  derive the asymptotic linearization of the NNRI estimator, which allows us to decompose the asymptotic variance of the NNRI estimator into two components  accounting for sampling and  matching. Based on the asymptotic variance formula, we propose an alternative variance estimator by approximating these  components by parametric or nonparametric approaches. We show the theoretical guarantees for the plug-in variance estimator.  
%We first study the statistical properties of the proposed variance estimator using asymptotic theory. 
The proposed variance estimator is easily applied to a variety of probability sampling designs and accounts for non-negligible sampling fractions. 

The rest of the paper is organized as follows. 
Section \ref{sec:basic} introduces the basic setup and the NNRI procedure in detail, including asymptotic properties. Section \ref{sec:results} derives variance estimation for the NNRI estimator.
In Section \ref{sec:application}, we apply the proposed variance estimator to empirical expenditures data for reference year 2018 from a subset of industries surveyed in the SAS. The studied data are typical of many business surveys, in that many units report a total but may not provide the associated detail items, especially when the item definitions are complex or the number of requested detail items is large \citep{willimack2000}.  The SAS uses a stratified simple random sample without replacement (SRS-WOR) design with high sampling rates in several strata. The empirical application highlights the differences between our proposed variance estimator from a naive variance estimator but does not provide insight into its statistical properties. Consequently, Section \ref{sec:simulation} investigates the  finite-sample performance of the proposed over repeated stratified SRS-WOR samples via a simulation study patterned off of \cite{andridge2020finding}. We close in Section \ref{sec:Conclusion} with some general observations along with directions for future research.

\section{Basic Setup}\label{sec:basic}
\subsection{Notation and assumptions}
Let ${{y}_{i}}=(y_{i1},\cdots,y_{iT})$ be the study variable of interest
and $x_{i}$ be the auxiliary variable. We assume that $x_{i}$ are
observed throughout the sample but $y_{i}$ are observed only for
the subset of the sample. Let $\delta_{i}=1$ if $y_{i}$ is observed
and $\delta_{i}=0$ otherwise. Let $I_i$ be the sampling indicator where $I_i=1$ if unit $i$ is selected, and otherwise $I_i=0$. Let $\mathcal{F}_N=\{(x_i,y_i,\delta_i):i=1,\cdots,N\}$ be a finite random sample from a superpopulation model $\zeta$ with known $N$. We make the following assumption for the missing data process. 
\begin{assumption}[Missing at random and positivity]\label{assump:MAR}
(i) The response indicator $\delta_i$ satisfies $P(\delta_i=1\mid x_i,y_i)=P(\delta_i=1\mid x_i)$, which can be denoted by $\pi(x_i)$, and (ii) $\pi(x_i)>\epsilon$ for a constant $\epsilon>0$ w.p. $1$.
\end{assumption}
Assumption \ref{assump:MAR} (i) states that the {response} indicator depends only on the observed $x$ but not on the outcome value $y$. Essentially, it assumes that the covariates contain all the information for the outcome that affects the probability of response, i.e., is missing at random in the population level (\citealp{rubin1976,berg2016imputation}). Assumption \ref{assump:MAR} (ii)  {indicates that all sampled units have a positive probability of responding given any possible value of $x$, in turn implying} that the support of the respondents {and the nonrespondents is} the same. {This assumption guarantees that \emph{all} donor values are plausible.} If the Assumption \ref{assump:MAR} (ii) is violated, our proposed estimator in Section \ref{subsec:NNRI} would no longer be asymptotically unbiased. Throughout, we assume this strong ignorability condition in Assumption \ref{assump:MAR} holds.

To motivate the NNRI estimator, we first consider the full response case using the Horvitz-Thompson (HT) estimator. Under full response, we can use 
\[
\widehat{T}_{y}=\sum_{i\in S}w_{i}y_{i}\]
to estimate $T_{y}=\sum_{i=1}^{N}y_{i}$, the population total of
$y_{i}$,  where $w_{i}$ is the sampling {(design)} weight  {computed as the inverse of the sampling inclusion probability} and $S$ is the index set of the sample with $|\mathcal{S}|=n$.

Let $E_p$ and $\text{var}_p$ be the expectation and variance with respect to the sampling mechanism; that is, $E_p(\cdot)=E(\cdot\mid \mathcal{F}_N)$ and $\text{var}_p(\cdot)=\text{var}(\cdot\mid \mathcal{F}_N)$. We assume a sequence of finite populations and samples in order to investigate the asymptotic properties as defined in \cite{fuller2011sampling}.
\begin{assumption}\label{assump:design}
The HT estimator of a population total given by $\hat{T}_{y}=\sum_{i\in \mathcal{S}}w_i y_i$ satisfies (i) $C_1\leq w_i n N^{-1}\leq C_2$; (ii) $\text{var}_p(N^{-1}\widehat{T}_{y})=O_p(n^{-1})$ and 
$\{\text{var}_p(\widehat{T}_{y})\}^{-1/2}(\widehat{T}_{y}-T_y)\mid \mathcal{F}_N \rightarrow \mathcal{N}(0,1)$ in distribution, as $n\rightarrow \infty$.
\end{assumption}
Assumption \ref{assump:design} is widely accepted in survey sampling {to allow for} valid inferential conclusion via asymptotic normality.

\subsection{Nearest neighbor ratio imputation estimator}\label{subsec:NNRI}
Nearest neighbor ratio imputation (NNRI) matches a donor to a recipient (nonrespondent), then multiplies the recipient's (available) total by the donated function $m(x_i)$ under the following assumption, assumed true for all $i\in \mathcal{S}$
\begin{equation}
 y_i = m(x_i) + e_i,\label{model}
\end{equation}
where $E_\zeta( e_i \mid x_i)=0$ and 
 $m(x_i ) = x_{i} R  (x_{i} ) $ for some smooth function $R( \cdot)$. Let $y_i^*$ be the imputed value of $y_i$ using NNRI as
$$ y_i^* = x_i {R}_{i(1)}, $$
where the ratio $R_i=\left(y_{i1},\cdots,y_{iT}\right)/x_i$ is only available from the responding units (donors), and $i(1)$ is the index of the nearest neighbor of unit $i$ within the {same imputation cell} where the nearest neighbor of $i$ satisfies 
\[
\mathcal{D}(x_{i},x_{i(1)})\le \mathcal{D}(x_{i},x_{j}),
\]
for all $j$ in the subsample of respondents, where $\mathcal{D}(\cdot,\cdot)$ is a suitable distance function (in this application, the absolute value of the distance).

Then, the imputed
estimator of $T_{y}$ is given by 
\begin{equation}
\widehat{T}_{y,I}=\sum_{i\in S}w_{i}\left\{ \delta_{i}y_{i}+(1-\delta_{i})y_{i}^{*}\right\} .\label{2}
\end{equation}
The goal is to estimate the variance of the imputed estimator in (\ref{2}).
If we define 
\[
d_{ij}=\left\{ \begin{array}{ll}
1 & \mbox{ if unit }i\mbox{ is used as a donor for unit }j,\\
0 & \mbox{ otherwise, }
\end{array}\right.
\]
then we can express 
\[
y_{i}^{*}=x_{i}{R}_{i(1)}=x_{i}\sum_{j\in\mathcal{S}}\delta_{j}d_{ji}{R}_j=\sum_{j\in \mathcal{S}}\delta_{j}d_{ji}\left(x_{i}/x_{j}\right)y_{j}.
\]
Thus, the imputed estimator in (\ref{2}) can be written as 
\begin{eqnarray}
\widehat{T}_{y,I} & = & \sum_{i\in S}w_{i}\left\{ \delta_{i}y_{i}+(1-\delta_{i})\sum_{j\in S}\delta_{j}d_{ji}\left(x_{i}/x_{j}\right)y_{j}\right\} \nonumber \\
 & = & \sum_{i\in S}\delta_{i}w_{i}(1+\kappa_{i})y_{i},\label{3}
\end{eqnarray}
where 
\[
\kappa_{i}=\sum_{j\in S}\frac{w_{j}x_{j}}{w_{i}x_{i}}(1-\delta_{j})d_{ij}.
\]
Note that $\kappa_{i}$ satisfies 
\begin{equation}
\sum_{i\in S}\delta_{i}w_{i}(1+\kappa_{i})x_{i}=\sum_{i\in S}w_{i}x_{i}.
\label{eq:kappa}
\end{equation}
To study the asymptotic properties of the NNRI estimator in (\ref{3}), we assume the following regularity condition holds. 
\begin{assumption}\label{ass:regular} 
(i) Let the matching variable $X$ be a random variable on a compact and convex support $\mathbb{X}$, with its density $f_X$ bounded and bounded away from zero. Suppose that $f_X$ is also differentiable in the interior of $\mathbb{X}$ with bounded derivatives; (ii) Let $R(x)$ be Lipschitz continuous in $x$, which means that $\exists C_3$ s.t. $\mid R(x_i)-R(x_j)\mid \leq C_3 \mid x_i-x_j\mid$, for any $i,j$.
\end{assumption}
Assumption \ref{ass:regular} (i) imposes a compact and convex support for the random variable $X$, which will be essential for studying the asymptotic properties of the NNRI estimator; Assumption \ref{ass:regular} (ii) restricts the ratio function $R_i$ to be smooth {in} $x_i$ \citep{abadie2006large,yang2020asymptotic}. Note that the underlying ratio model in (\ref{model}) is a special case of the general models $m(\cdot)$ that clearly satisfies the Lipschitz condition since the Lipshitz condition on $R(x)$ implies that on $m(x)$.

The following lemma describes the key asymptotic property  %induced by (\ref{eq:kappa}) 
with the proof deferred to the Supplementary Material.
\begin{lemma}
\label{lem:asym_equal}
Under Assumptions \ref{assump:MAR}--\ref{ass:regular}, we have
\begin{equation}
   \sum_{i\in \mathcal{S}}\delta_i w_i(1+\kappa_i)x_i R(x_i)=
\sum_{i\in\mathcal{S}}w_ix_i R(x_i)+O_p(n^{-1}N). \label{eq:kappa2}
\end{equation}
Combining with \eqref{eq:kappa2}, we have 
\begin{equation}
n^{1/2}N^{-1} \widehat{T}_{y,I}=n^{1/2}N^{-1} \sum_{i\in S}w_{i}\left[ x_{i}{R}(x_i)+\delta_{i}(1+\kappa_{i})\{y_{i}-x_{i}{R}(x_i)\}\right]+o_p(1). \label{4}
\end{equation}
\end{lemma}
Considering that $E(y_{i}\mid x_{i})=x_{i}R(x_i)$ in (\ref{model}),
we can express (\ref{4}) as 
\begin{equation}
n^{1/2}N^{-1} \widehat{T}_{y,I}=n^{1/2}N^{-1}\sum_{i\in S}w_{i}\left\{ {m}_{i}+\delta_{i}(1+\kappa_{i}){e}_{i}\right\}+o_p(1), \label{5}
\end{equation}
where ${m}_i=x_i{R}(x_i)$ and ${e}_{i}=y_{i}-{m}_{i}$. This decomposition assures that the first component $m_i$ is uncorrelated with the second component $e_i$ under the conditional argument of $x_i$. 
The decomposition leads to the asymptotic distribution of the NNRI estimator as follows. 
\begin{theorem}\label{theorem:normal} Under Assumptions \ref{assump:MAR}--\ref{ass:regular}, suppose the ratio model in (\ref{model}) holds true and define $\sigma^2_e(x_i)=\text{var}(e_i\mid x_i)=E(e_i^2\mid x_i)$. Then, $n^{1/2}N^{-1}(\widehat{T}_{y,I}-T_{y})
\rightarrow \mathcal{N}(0,V_y)$ in distribution as $n\rightarrow \infty$, where
$$
V_y=V^m+V^e,
$$
with 
$$
    V^m=\lim_{n\rightarrow \infty}\frac{n}{N^2}E\left\{\text{var}_p
    \left(\sum_{i\in S}w_{i}m_{i}-\sum_{i=1}^{N}m_{i}\right)\right\},$$and
    \begin{equation}V^e= 
 \lim_{n\rightarrow \infty}\frac{n}{N^2}E_{}\left[\sum_{i=1}^{N}\left\{ I_{i}w_{i}\delta_{i}(1+\kappa_{i})-1\right\}^2 \sigma_e^2(x_i)\right]. \label{ve1}
\end{equation}
In particular, if $n/N=o(1)$, then $V^e$ reduces to
 \begin{equation}
V^{e}=\lim_{n\rightarrow \infty}\frac{n}{N^2}\sum_{i=1}^{N}E_{}\left[ I_{i}\delta_{i}\left\{ w_{i}(1+\kappa_{i})\right\} ^{2}\sigma^2_e(x_i)\right].\label{ve2}
 \end{equation}
\end{theorem}

The detailed proof of Theorem \ref{theorem:normal} is presented in the Supplementary Material. 
The results in Theorem \ref{theorem:normal} are obtained by taking {the reverse sampling} arguments following \cite{shao1999variance} and \cite{kim2006}, so that the sample-response path begin with a census with nonrespondents from which a sample is selected.  In the {reverse sampling} framework, the outer expectation (denoted $E$) is taken with respect to {the superpopulation model for} the census with nonrespondents, with inner expectations and variances {with respect to the sampling design} conditional on $(\delta_1, \cdots, \delta_N)$. The component $V^m$ is the variance due to the sampling design. The other  component, $V^e$, constitutes the error variance added due to the deterministic ratio imputation model via the NNRI. Variance formula (\ref{ve1}) requires access to population $x_i$'s whereas variance formula (\ref{ve2}) does not provided that the overall sampling fraction is negligible. 

\begin{comment}\label{rem:NNI_NNRI}
NNI v.s. NNRI
\textcolor{blue}{I am having a hard time proving this remark analytically, and I ran a little simulation, the results is not in our favor; see below for the MC standard error of NNRI and NNI
\begin{tabular}{cc}
\hline
 est.NNRI &  est.NNI\\
1217.7456 &1216.1651\\
768.9328  &768.9963\\
364.9642  &365.0880\\
179.9780  &179.4599\\
150.4431  &149.3954\\
\hline
\end{tabular}
}
\end{comment}

\section{Variance Estimation}\label{sec:results}

\cite{yang2019nearest,yang2020asymptotic} describe asymptotically unbiased variance estimators that fully account for the error term ($V^m$ and $V^e$) presented in Section \ref{subsec:NNRI}, under predictive mean matching imputation. Since NNRI is a special case of predictive mean matching, we modify their variance estimator to obtain asymptotically unbiased estimators i.e., $E(\widehat{V}^m+\widehat{V}^e)=\text{var}\{n^{1/2}N^{-1}(\widehat{T}_{y,I}-T_y)\}$.

Both the empirical application presented in Section \ref{sec:application} and the simulation study presented in Section \ref{sec:simulation} employ stratified SRS-WOR designs that include a certainty stratum (all units sampled with probability $=1$) and at least one sampling strata with a large sampling rate (greater than $0.50$). The sample design is highly characteristic of business surveys, which are sampled from highly skewed populations. The presented applications use approximate sampling variance (design based) estimators for $\widehat{V}_m$ to easily incorporate the non-negligible sampling fractions, although we also introduce a general replication variance estimator for use when sampling fractions are negligible.

\subsection{Estimation of \texorpdfstring{${m}_i$}{}}\label{subsec:mhat}

Assumption 3 requires a smooth estimator of $R(x)$ that can be used for \emph{all} sampled units to estimate ${m}_i$ as $x_i R(x_i)$. However, NNRI is not a smooth imputation procedure. We approximate a smooth ratio function $R(x)$ in \eqref{4} with a plug-in estimator ($\widehat{R}(x_i)$), considering
\begin{enumerate} 
    \item PARAM1 Parametric ratio estimator with $\widehat{R}(x_i) = \widehat{\beta} =\sum_{i \in S} w_{i} \delta_{i}y_{i}/\sum_{i \in S} w_{i} \delta_{i} x_{i}$, the  B.L.U.E. of the weighted simple linear no-intercept regression model $y_{i}w_{i}^{-1/2} =\beta x_i w_{i}^{-1/2} + e_{i} w_{i}^{-1/2}$, $e_{i} \sim (0, x_i\sigma^2)$. This is the PAR1 estimator utilized in \cite{beaumont2009variance};
    \item PARAM2 Parametric ratio estimator with $\widehat{R}(x_i)$ = $\widehat{\beta}_h$, where $\widehat{\beta}_h$ is estimated separately within each sampling stratum $h$ by fitting the regression model from (a); and
    \item NONPARAM Generalized Additive Model (GAM) estimator \citep{hastie1990generalized} approximating the unknown smooth function of $R(x_i)$ with multinomial link functions.
\end{enumerate}
The first two estimators are frequently employed in the survey research methods literature (e.g., \citet{magee1998}, among others). Of course, these estimators require a very specific relationship between the independent and auxiliary variable, and this strong association is less likely for rarely-reported independent variables. Furthermore, the parametric approximations develop \emph{separate} regression models for each detail item component in $y_i$. As a nonparametric alternative approach, we {appeal to the GAM for a more flexible representation. In short, GAMs can be considered as fitting several spline smoothers to approximate the \emph{unknown} smooth function, in this case the ratio function ${R}(x)$.  By applying the order-$n$ basis expansions of $R(x)$ under the multinomial link function, one could observe that for $i=1,\cdots, n$}
\begin{align}
&\eta^{(t)}(x_i)=
    \sum_{k=1}^{n}
    \beta_k^{(t)}b_k^{(t)}(x_i), t = 1,\cdots , T-1;
    \eta^{(T)}(x_i)=1,\\
    &\widehat{R}(x_i)
 =\left\{
 \frac{\exp\{\eta^{(1)}(x_i)\}}{\sum_{t=1}^T \exp\{\eta^{(t)}(x_i)\}},
 \frac{\exp\{\eta^{(2)}(x_i)\}}{\sum_{t=1}^T \exp\{\eta^{(t)}(x_i)\}},\cdots,
 \frac{\exp\{\eta^{(T)}(x_i)\}}{\sum_{t=1}^T \exp\{\eta^{(t)}(x_i)\}}
 \right\}.
 \label{ratio:nonparam}
\end{align}
The $\{{\beta}_k^{(t)}\}_{k=1}^{n},t=1,2,\cdots,T-1$ are the regression coefficients for the $t-$th detail item and $\{b_k^{(t)}(x_i)\}_{k=1}^{n}, t=1,2,\cdots,T-1$ are the known $n$ basis functions ({e.g., splines, radial functions, etc.}) associated with the $n$ sample points, which are usually {assumed to have} good approximation theoretical properties. To {alleviate the overfitting issue, a model penalty can be imposed} during the model fitting to reduce the number of basis functions from $n$ to $K$ \citep{wood2003thin,wood2016smoothing}. The penalized objection function {is} specified as
$$
\min_{\boldsymbol{\beta}}\left\{\sum_{i=1}^n\delta_i\|x_i{{\widehat{R}(x_i)}}-y_i\|^2+\lambda
{J\{\widehat{R}(x)\}}\right\},
$$
where the first term measures the closeness of our fitted functions while the second term {$J\{\widehat{R}(x)\}$} penalizes the wiggliness of the function associated with the tuning parameter $\lambda$, which can be obtained by the cross validation technique. Here, we adopt the wiggliness penalty functional from \cite{wood2003thin} illustrated in  Example \ref{eg:tp}.
\begin{example}
\label{eg:tp}
Let $\widehat{R}(x)\in\mathcal{H}$, where $\mathcal{H}$ is an arbitrary reproducing kernel Hilbert space. Begin with a simple objective function:
\begin{equation}
\label{eq:B-spline}
    \min_{\boldsymbol{\beta}}\sum_{i=1}^n\delta_i\|x_i\widehat{R}(x)-y\|^2+\lambda \int_a^b \widehat{R}^{''}(x)^\intercal \widehat{R}^{''}(x) dx
\end{equation}
with arbitrary value $a$ and $b$ as long as they cover the variable in question \citep{hastie1990generalized}. The resulting solution $\widehat{R}^*(x)$ can be considered as a type of the \emph{basis spline} or B-spline (See Figure \ref{fig:GAM:toy} for an illustrative example). However, its estimation requires $O(n^3)$ operations in the univariate case, which is computational infeasible for the large-scale of datasets. One approach to tackle this problem is to employ the \emph{regression splines}, from which an optimal approximation of $\widehat{R}(x)$ can be produced via choosing the truncated bases with lower ranks $K^{(t)}$ for $t=1,\cdots, T-1$ \citep{wood2003thin}.
\begin{figure}[htbp]
    \centering
    \includegraphics[width=.8\linewidth]{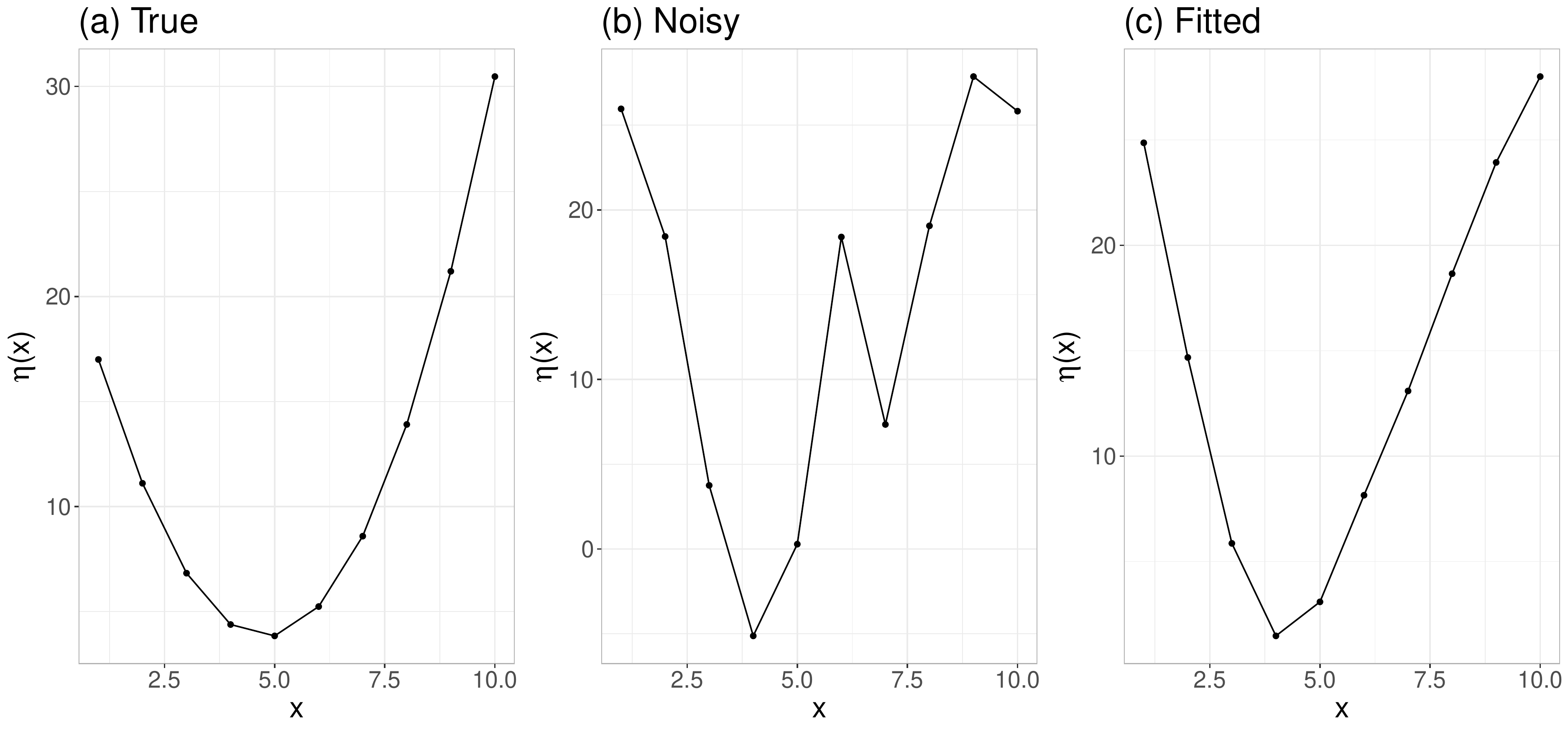}
    \caption{An illustrative example of the nonparametric B-spline, where $x$ is a scalar variable ranging from $1$ to $10$, with $10$ observed sample points as $1,2,\cdots, 10$. Leftmost (a) is the underlying true values of $\eta(x)=\sqrt{|x|}+|x-5|^2+\log(|x|)$; Middle (b) is the observed values of $\eta(x)$ corrupted by the additive error term $e\sim \mathcal{N}(0,10^2)$ ; Rightmost (c) is the fitted values of $\eta(x)$ by implementing the B-spline approach based on (\ref{eq:B-spline}).}
    \label{fig:GAM:toy}
\end{figure}
\end{example}

\subsection{Estimation of \texorpdfstring{${V}^m$}{}}
Using the pseudo observations $\widehat{m}_i$ with postulated parametric or nonparametric models of $R(x)$ in Section \ref{subsec:mhat}, a standard design-based estimator under complete response is given by
\begin{equation}
\label{11} 
\widehat{V}^m = 
\sum_{i\in\mathcal{S}}
\sum_{j\in\mathcal{S}}
\Omega_{ij}
\widehat{m}_i
\widehat{m}_j,
\end{equation}
where $\Omega_{ij}$ accounts for various sampling designs. For example, under simple random sampling, the variance expression of $V_m$ simplifies to $(1-n/N)(n-1)^{-1}
\sum_{i \in \mathcal{S}}(\widehat{m}_i-\bar{{m}})^2
$ with $\bar{{m}}=1/n\sum_{i\in \mathcal{S}}\widehat{m}_i$. With a stratified SRS-WOR design as in Sections \ref{sec:application} and \ref{sec:simulation}, the same formula is used to obtain $\widehat{V}^{mh}$ independently in each stratum, then aggregated ($\widehat{V}^m=\sum_{h}\widehat{V}^{mh}$).

For sampling designs with negligible sampling fractions, the ease of a replication variance estimator \citep{wolter2007introduction} may be preferable for the NNRI estimator. The general replicate variance estimator when $y_i$ is observed throughout the sample is  
\begin{equation}
\widehat{V}_{\text{rep}}(\widehat{T}_y)=\frac{n}{N^2}\sum_{k=1}^{L}c_{k}\left(\widehat{T}^{(k)}_y-\widehat{T}_y\right)^{2},\label{8}
\end{equation}
where $c_k$ is the $k$th replication factor, and $
\widehat{T}_y^{(k)}=\sum_{i\in \mathcal{S}}w_{i}^{(k)} y_i$ in which $w_{i}^{(k)}$ is the $k$-th replicate weight for unit $i$, using a replication method that appropriately accounts for the complex sampling design [Note: for a stratified SRS-WOR sample with non-negligible sampling fractions, the $c_k$ should be modified to include the finite population correction factors ($1 - n_h/N_h$)]. The replicates are constructed such that $E\{\widehat{V}_{\text{rep}}(\widehat{T}_y)\}=\text{var}\{n^{1/2}N^{-1}(\widehat{T}_y-T_y)\}\{1+o(1)\}$. We illustrate replicate weights through the following example. 
\begin{example}
Suppose that the probability sample is obtained by a single-stage design where each unit $i$ has a sampling weight $w_i$, the delete-1-jackknife method yields an unbiased estimate of the sampling variance under complete response. Therefore, $L=n, c_k=(n-1)/n$, , and $w_i^{(k)}=nw_i/(n-1)$ if $i\neq k$, and $w_k^{(k)}=0$ if $i$ = $k$.
\end{example}

\subsection{Estimation of \texorpdfstring{${{V}^e}$}{}}\label{subsec:Ve}
We now discuss estimation of the $V^e$ term. If an asymptotically unbiased estimator of $\sigma_e^2 (x_i)$ is available, then we may use 
\begin{equation}
    \widehat{V}^e=
\frac{n}{N^2}
\sum_{i \in \mathcal{S}}
\{
w_i^2\delta_i (1+\kappa_i)^2+w_i-
2w_i\delta_i(1+\kappa_i)
\}
\widehat{\sigma}_e^2(x_i).
\label{eq:V_e}
\end{equation}
If additionally, we assume the Lipschitz continuity of $\sigma_e^2(x_i)$ in $x$, a similar result to Lemma \ref{lem:asym_equal} can be obtained
\begin{equation}
   \sum_{i\in\mathcal{S}}\delta_i w_i (1+\kappa_i)\sigma_e^2(x_i) = 
\sum_{i\in\mathcal{S}}w_i \sigma_e^2(x_i) + O_p(n^{-1}N). 
\label{eq:V_e_equal}
\end{equation}
Substituting (\ref{eq:V_e_equal}) back into (\ref{ve1}) yields 
\begin{equation}
    {V}^e = \frac{n}{N^2}
    \sum_{i\in \mathcal{S}}
    \{w_i^2\delta_i(1+\kappa_i)^2 - w_i\delta_i(1+\kappa_i)\}{\sigma}_e^2(x_i)
\end{equation}
Now, we can directly use the residuals $\widehat{e}_i$ obtained from the modeled values i.e., $\widehat{e}_i = y_i - \widehat{m}_i$ to estimate $\sigma_e^2(x_i)$, where $\widehat{m}_i$ is obtained using the PARAM1, PARAM2, or NONPARAM models presented in Section \ref{subsec:mhat}.  In particular, we have
$$
\widehat{V}^e = \frac{n}{N^2}
    \sum_{i\in \mathcal{S}}
    \{w_i^2\delta_i(1+\kappa_i)^2 - w_i\delta_i(1+\kappa_i)\}\widehat{e}_i^2.
$$
Alternatively, we can model the variation of the residuals to estimate ${\sigma}_e^{2}(x_{i})$ in (\ref{eq:V_e}). We consider three approaches:
\begin{enumerate}
    \item PARAM1(M) Plug-in variance estimator as $\widehat{\sigma}_e^{2}(x_{i})=x_{i}\widehat{\beta}(1- \widehat{\beta})$, specified by the multinomial distribution of $y_i$, where $\widehat{\beta}$ is the PARAM1 estimator from Section \ref{subsec:mhat}.
    \item PARAM2(M) Parametric linear estimator of $\widehat{\sigma}_e^{2}(x_{i})=\widehat{\alpha}_{0,h}+ \widehat{\alpha}_{1,h} x_i$ obtained by the OLS regression of $e_i^2= (y_i - \widehat{\beta}_h x_i)^2 = \alpha_{0,h}+\alpha_{1,h} x_i$ {for $\delta_i=1$ within each strata , where $\widehat{\beta}_h$ is PARAM2 estimator from Section \ref{subsec:mhat}}.
    \item NONPARAM(M) Nonparametric estimator of $\widehat{\sigma}_e^{2}(x_{i}) = \sum_{k=1}^K \widehat{\beta}_k b_k(x_i)$ obtained by fitting a GAM of $e_i^2=\delta_i(y_i-\widehat{m}_i)^2 = \sum_{k=1}^K \beta_k b_k(x_i)$ for all strata , where $\widehat{m}_i$ is obtained using the NONPARAM estimator described in Section \ref{subsec:mhat}.
\end{enumerate}

Combining the $V^m$ components obtained with the proposed estimators of $m_i$ described in Section \ref{subsec:mhat} and the  $V^e$ components yields six candidate variance estimators. 

\section{Empirical Application (Service Annual Survey)}\label{sec:application}

In this section, we apply the proposed variance estimator to empirical data from the 2018 collection of the Service Annual Survey (SAS). Conducted by the U.S. Census Bureau, the SAS is a mandatory survey of approximately 78,000 employer businesses (companies) having one or more establishments located in the U.S. that provide services to individuals, businesses, and governments.  The SAS collects aggregate and detailed revenues and expenses, e-commerce, exports and inventories data from a stratified sample of business firms with paid employees in selected industries in the services sector. As mentioned in Section 1, the key items collected by SAS are total revenue and total expenses; detailed breakdowns of these two totals items are requested from all sampled firms.  The revenue detail items vary by industry within sector. Expense detail items, however, are primarily the same for all sectors, with an occasional additional expense detail or two collected for select industries.  Complete information on the SAS methodology is available at \url{https://www.census.gov/programs-surveys/sas/technical-documentation/methodology.html}.

SAS uses imputation to account for unit and item nonresponse, relying heavily on the ratio imputation models presented in Section \ref{subsec:mhat} (specifically, PARAM1 and PARAM2). For total receipts and total expenses, the independent variables are either a highly correlated data item from the same reference period or historic data for the same item. The model for detail items use the corresponding totals item (reference period only). \citet{thompson2013} evaluated these imputation models in two of the sectors included in the SAS, explicitly fitting weighted no-intercept linear regression models within industry using respondent data to assess model fit (i.e., the PARAM1 method). For the totals, the ratio imputation models were appropriate, with adjusted-$R^2$ consistently above 95\%. Given such strong predictors, the nonresponse adjustment is robust to the assumed response mechanism and should decrease the variance \citep{little2005var}.  However, the results for the details items were far less convincing, with adjusted-$R^2$ values often well-below 75-percent and non-significant slopes ($\alpha = 0.10$). The weak predictive power of this ratio imputation approach is exacerbated in the 2020 data collection year, as many businesses were closed or had business limited due to the COVID-19 pandemic. Not surprisingly, the SAS program managers were interested in exploring other imputation methods for these detail items, which are historically less frequently and reliably reported than their associated totals and whose imputed values are generally more difficult to independently validate. 

The empirical application is restricted to five industries, collectively representing a cross-section of the survey's \emph{expenditures} data collections. We chose a subset of industries from a candidate list provided by subject matter experts, requiring a minimum of three sampled units per strata in addition to validating that the NNRI model assumptions appeared to hold.  Consequently, these industries are \emph{not} representative of the larger survey. The input data for this application study consists of the sampled companies in the selected industries that tabulated a \emph{positive} (non-zero) total expenses value; companies reporting zero-valued expenses were dropped. Furthermore, the SAS uses industry-average ratio imputation (not NNRI) for missing and invalid expenses items and implements a \text{naïve} random group variance estimator for all item. For these reasons, our estimates and variance estimates differ from the official published values.

 Tables \ref{tab:Rates} and \ref{tab:Design} describes selected features of the study industries. The sample sizes and response rates are rounded to comply with internal regulations. The unweighted response rates presented in Table \ref{tab:Rates} represent the proportion of sampled units that provided \emph{a complete donor record} and do not correspond to the official response rates.  On inspection, the response patterns displayed in Table \ref{tab:Rates} appear to be atypical of business surveys, in that generally the larger businesses (e.g., the certainty companies) respond at higher rates. However for NNRI, \emph{all} detail items are imputed if either (1) the company was a nonrespondent or (2) the reported set of expenditures details did not add to the total, thus reducing the total number of respondents.  

\begin{table}
\caption{
\label{tab:Rates}Number of expenditures detail items (details) and unweighted response rates of expenditures details (in percentages) by certainty and noncertainty status for SAS study industries. Businesses that are included with certainty are sampled with probability = 1; noncertainty businesses are sampled with probability less than 1}
\centering
\resizebox{\textwidth}{!}{%
\fbox{%
\begin{tabular}{*{6}{c}}
\multirow{2}{*}{Industry} & \multirow{ 2}{*}{Description} & Details   & \multicolumn{3}{c} {Response Rates (in Percentages)} \\
& & $y_{i}$ &Total&  Certainty & Noncertainty  \\
\hline
221122 &	Electric Power Distribution	            & 7	& 75	& 65	& 95 \\
517210 &	Wireless Telecommunications Carriers 	& 9	& 30	& 60	& 30 \\
541211 &	Offices of Certified Public Accountants	& 7	& 80	& 75	& 85 \\
621410 &	Family Planning Centers	                & 9	& 85	& 80	& 90 \\
713110 &	Amusement and Theme Parks	            & 7	& 60	& 60	& 60
\end{tabular}%
}
}
\end{table}

\begin{figure}
    \centering
    \includegraphics[width=.8\linewidth]{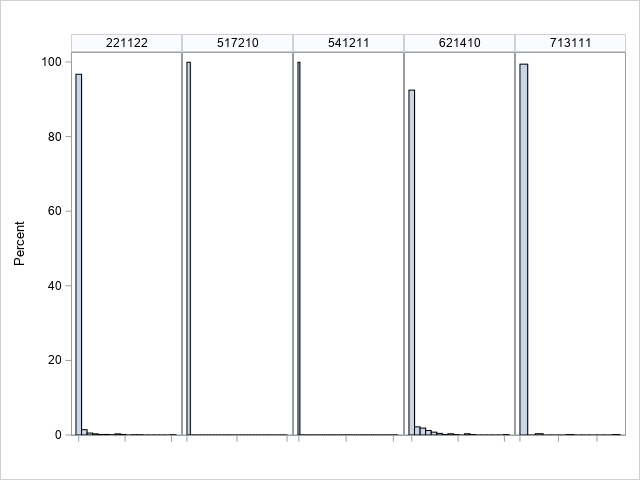}
    \caption{Histograms of total expenses within industry for the studied industries. Data source: Service Annual Survey (2018, U.S. Census Bureau.}
    \label{emp_histograms}
\end{figure}

Figure \ref{emp_histograms} presents histograms of total expenses within study industry. Data values are suppressed for confidentiality protection. Typical of many business surveys, all five distributions are highly positively skewed, and the largest businesses are sampled with probability = 1 (certainty). These  distributions of the total expenses variable are well-approximated by lognormal distributions in all industries. Notice that the variance of the approximate lognormal distribution is small in three industries. In three industries (517210, 541210, and 713110), so that the compact support requirement in Assumption \ref{ass:regular} is approximately true. Unfortunately, conformance to this requirement is unlikely for the other two study industries, although the convex support assumption should hold for all five industries.

Table \ref{tab:Design} provides sample design characteristics of the study industries. The overall sampling rate ($n/N$) is non-negligible in four of the five study industries. All industries include a certainty strata (all units included with probability $= 1$), which is excluded from the range provided in Table \ref{tab:Design} but is included in the computation of the overall sampling rate.  The sampling rates for the noncertainty industry strata vary greatly, with at least one stratum in each industry having a non-negligible sampling rate. 

\begin{table}
\caption{
\label{tab:Design} Sample design characteristics of SAS study industries. The range of strata sampling rates excludes the certainty stratum}
\centering
\resizebox{\textwidth}{!}{%
\fbox{%
\begin{tabular}{*{7}{c}}
\multirow{ 2}{*}{Industry}  & Overall   & \multicolumn{2}{c} {Sample Size ($n$)} & Number of & \multicolumn{2}{l} \textbf{Strata Sampling Rates ($n_{h}/N_{h})$} \\
& $n/N$ & Certainty & Noncertainty & Strata & Minimum & Maximum \\
\hline
221122  & 0.0785 & 80 & 30 & 7 & 0.0061 & 0.3448  \\
517210  & 0.0620 & 60 & 350 & 15 & 0.0140 & 0.5000 \\
541211  & 0.0035 & 40 & 150 & 13 & 0.0020 & 0.0894  \\
621410  & 0.1055 & 50 & 50 & 7 & 0.0122 & 0.4468  \\
713110  & 0.0709 & 40 & 30 & 4 & 0.0160 & 0.1125 \\ 
\end{tabular}%
}}
\end{table}

Table \ref{tab:emp_ratio} provides ratios of the NNRI detail item totals to their corresponding total expenses values (denoted $\hat R_y$) and ratios of the NNRI detail items variance estimates computed with our proposed variance estimators to their corresponding $\text{naïve}$ variance counterparts (denoted $R_{METHOD}$).  The  $\hat R_y$ estimates are presented without error bounds and are provided to illustrate the prevalence of a reported detail item in an industry. Since the $\text{naïve}$ estimates do not account for variance caused by nonresponse and imputation, we expect the $\text{naïve}$ variance estimates to be smaller than those obtained by the proposed variance estimates in general, although ratios close to 1 could indicate an extremely effective hot deck prediction.  Variance ratios greater than $1,000$ are indicated by an ``XXX.''

\footnotesize
\captionsetup{width=\textwidth}
\begin{longtable}{llccccccccc}
\caption{
\label{tab:emp_ratio} Ratios of NNRI detail item totals to total expenses (line $1$ in each set of industry results) and ratios of proposed variance estimators to corresponding naive variance estimates within industry (lines $2 -7$ in each set of industry results). The detail item proportions will not necessarily add to $1$ due to rounding. Entries with ``XXX'' indicate ratio values greater than 1000. For all industries, the detail items are Y1 = gross annual payroll; Y2 = fringe benefits; Y3 = temporary staff payroll; Y4 = expensed software; Y5 = depreciation costs; Y6 = expensed equipment; and Y7 = other expenses. In industry 517210, Y8 = access charges and Y9 = universal service and similar charges. In industry 621410, Y8 = professional liability insurance and Y9 = medical supply costs. Data Source: Service Annual Survey (2018), U.S. Census Bureau}\\
\toprule
  \centering
    Industry & Variance &  Y1   &  Y2   &  Y3   &  Y4   &  Y5   &  Y6   &  Y7   &  Y8   &  Y9 \\
    \midrule
    \multirow{7}[2]{*}{221122} &  $\widehat{R}_y$ & 0.1   & 0.0     & 0.0   & 0.0   & 0.2   & 0.0   & 0.6   &       &  \\
          & ${R}_\text{PARAM1}$ & 6.2   & 2.9   &  XXX  & 14.4  & 25.9  & 7.7   & 1.6   &       &  \\
          & ${R}_\text{PARAM1 (M)}$ & 1.5   & 0.5   & 82.0  & 0.1   & 4.0   & 0.1   & 0.4   &       &  \\
          & ${R}_\text{PARAM2}$ & 16.8  & 2.9   &  XXX  & 18.0  & 19.7  & 2.7   & 0.8   &       &  \\
          & ${R}_\text{PARAM2 (M)}$ & 16.7  & 6.1   &  XXX  & 91.4  & 47.4  & 494.4 & 40.2  &       &  \\
          & ${R}_\text{NONPARAM}$ & 8.6   & 3.8   &  XXX  & 2.0   & 9.3   & 2.0   & 0.7   &       &  \\
          & ${R}_\text{NONPARAM (M)}$ & 8.6   & 3.8   &  XXX  & 2.0   & 9.3   & 2.0   & 0.7   &       &  \\
    \midrule
    \multirow{7}[2]{*}{517210} & $\widehat{R}_y$ & 0.1   & 0.0   & 0.0   & 0.1   & 0.2   & 0.3   & 0.3   & 0.0  & 0.0 \\
          & ${R}_\text{PARAM1}$ &  XXX  & 677.6 &  XXX  & 1.3   &  XXX  &  XXX  & 1.3   & 9.6   & 0.5 \\
          & ${R}_\text{PARAM1 (M)}$ & 0.1   & 0.2   & 0.5   & 1.0   & 2.5   & 2.6   & 0.3   & 0.0   & 0.4 \\
          &  ${R}_\text{PARAM2}$ & 4.3   & 2.5   & 4.5   & 6.3   & 13.7  & 18.8  & 1.4   & 2.9   & 5.1 \\
          & ${R}_\text{PARAM2 (M)}$ & 3.2   & 171.4 & 292.1 & 3.3   & 7.6   & 612.1 & 243   & 15.9  & 787.0 \\
          & ${R}_\text{NONPARAM}$ & 2.5   & 2.7   & 2.5   & 1.2   & 10.1  & 4.2   & 2.2   & 2.3   & 2.8 \\
          & ${R}_\text{NONPARAM (M)}$ & 3.5   & 2.7   & 3.3   & 4.4   & 8.1   & 13    & 1.7   & 2.9   & 4.2 \\
    \midrule
    \multirow{7}[2]{*}{541211} &  $\widehat{R}_y$  & 0.5   & 0.1   & 0.0   & 0.0   & 0.0   & 0.0   & 0.3   &       &  \\
          & ${R}_\text{PARAM1}$ & 1.0   & 1.9   & 9.9   & 1.2   & 2.0   & 1.4   & 2.6   &       &  \\
          & ${R}_\text{PARAM1 (M)}$ & 0.7   & 0.5   & 0.3   & 0.2   & 0.2   & 0.0   & 1.0   &       &  \\
          &  ${R}_\text{PARAM2}$ & 0.9   & 1.5   & 2.2   & 1.1   & 1.7   & 0.9   & 2.1   &       &  \\
          & ${R}_\text{PARAM2 (M)}$  & 1.1   & 124.4 & 15.5  & 1.3   & 89.3  &  XXX  & 6.8   &       &  \\
          & ${R}_\text{NONPARAM}$ & 1.0   & 1.2   & 1.3   & 1.1   & 1.2   & 0.8   & 1.9   &       &  \\
          & ${R}_\text{NONPARAM (M)}$ & 1.0   & 1.4   & 1.7   & 1.0   & 1.6   & 0.8   & 1.3   &       &  \\
    \midrule
    \multirow{7}[2]{*}{621410} &  $\widehat{R}_y$ & 0.5   & 0.1   & 0.0   & 0.0   & 0.0   & 0.0   & 0.3   & 0.0   & 0.1 \\
          & ${R}_\text{PARAM1}$ & 0.9   & 2.0   & 4.7   & 3.5   & 3.8   & 1.4   & 1.5   & 1.7   & 1.5 \\
          & ${R}_\text{PARAM1 (M)}$ & 0.5   & 1.5   & 0.4   & 0.1   & 0.8   & 0.1   & 0.3   & 0.1   & 0.7 \\
          & ${R}_\text{PARAM2}$ & 1.0   & 7.0   & 4.5   & 1.4   & 4.6   & 1.4   & 1.0   & 1.1   & 4.5 \\
          &${R}_\text{PARAM2 (M)}$  & 19.9  & 7.1   & 5.1   & 52.7  & 7.2   & 87.3  & 107.8 & 158.6 & 19.2 \\
          &  ${R}_\text{NONPARAM}$ & 1.0   & 6.2   & 3.5   & 1.1   & 5.1   & 1.4   & 1.0   & 1.3   & 3.6 \\
          & ${R}_\text{NONPARAM (M)}$  & 0.8   & 2.5   & 3.6   & 1.2   & 1.9   & 1.5   & 1.2   & 1.0   & 2.1 \\
    \midrule
    \multirow{7}[2]{*}{713110} & $\widehat{R}_y$ & 0.3   & 0.1   & 0.0   & 0.0   & 0.2   & 0.0   & 0.4   &       &  \\
          & ${R}_\text{PARAM1}$ &  XXX  & 371   & 38.7  & 269.3 & 100.2 &  XXX  & 85.9  &       &  \\
          & ${R}_\text{PARAM1 (M)}$ & 0.4   & 1.3   & 0.1   & 0.7   & 0.5   & 0.0   & 0.5   &       &  \\
          &  ${R}_\text{PARAM2}$ & 6.0   & 31.2  & 5.6   & 344.1 & 2.4   & 7.4   & 5.2   &       &  \\
          & ${R}_\text{PARAM2 (M)}$ & 1.0   & 14.1  & 3.4   &  XXX  & 7.6   & 51.7  & 2.1   &       &  \\
          & ${R}_\text{NONPARAM}$ & 1.5   & 2.5   & 0.0   & 58.8  & 2.7   & 3.6   & 2.1   &       &  \\
          & ${R}_\text{NONPARAM (M)}$ & 6.6   & 30.1  & 5.6   & 343.3 & 2.2   & 7.2   & 5.3   &       &  \\
          \bottomrule
\end{longtable}%
\normalsize

The parametric ratio estimator employed by the $\hat V_\text{PARAM1}$ is clearly problematic. The variance estimates obtained with the $\hat V_\text{PARAM1}$ for the rarely reported detail items (values less than 10\%)  are often considerably larger than the counterparts obtained with non-\text{naïve} variance estimates. However, using modeled residuals for $\sigma^2_e$ appears to underestimate the variances, as evidenced by frequency of variance ratios with values less than one. Similarly the variance estimates obtained using the modeled residuals with the PARAM2 model ($\hat V_\text{PARAM2(M)}$)  tend to be much larger than those obtained with any other considered variance estimator, providing evidence of a poor model fit for $\sigma^2_e$. 

Table \ref{tab:cv} presents the coefficients of variation (c.v's) of each item for each considered variance estimator in percentages. At the 95\% confidence level, a total with an associated c.v. greater than $51\%$ ($= 1/1.96$) is not significantly different from zero. Thus, the c.v.'s provide a measure of the practical impact of the variance estimators on inference. 

\footnotesize
\captionsetup{width=\textwidth}
\begin{longtable}{llcccccccccc}
\caption{\label{tab:cv}Coefficients of Variation (c.v.) for NNRI Detail Items (HT Totals) in Percentages. Data Source: Service Annual Survey (2018), U.S. Census Bureau}\\
\toprule
\centering
{Industry} & {Variance} & Y1 & Y2 & Y3 & Y4 & Y5 & Y6 & Y7 & Y8 & Y9 \\ \hline
 & $\hat V_\text{Naïve}$ & 2.2 & 3.9 & 0.3 & 7.1 & 1.3 & 10.9 & 4.2 & \multicolumn{2}{c}{} \\ 
 & $\hat V_\text{PARAM1}$ & 5.4	& 6.6 & 34.1 & 27.0 & 6.8 & 30.2 & 5.4  & \multicolumn{2}{c}{} \\ 
 & $\hat V_\text{PARAM1(M)}$ & 2.6 & 2.7 & 2.6 & 2.6 & 2.7 & 3.1 & 2.8 & \multicolumn{2}{c}{} \\ 
 &  $\hat V_\text{PARAM2}$ & 8.9 & 6.6 & 27.2 &	30.2 & 6.0 & 17.9 & 3.7  & \multicolumn{2}{c}{} \\ 
 &  $\hat V_\text{PARAM2(M)}$ & 8.9 & 9.6 & 106.3 & 68.0 & 9.3 & 242.0 & 26.9 & \multicolumn{2}{c}{} \\ 
 &  $\hat V_\text{NONPARAM}$ & 6.4 & 7.6 & 14.7 & 10.1 & 4.1	& 15.6 & 3.6 & \multicolumn{2}{c}{} \\ 
\multirow{-7}{*}{221122} & $\hat V_\text{NONPARAM(M)}$ & 8.7 & 6.9 & 27.6 & 29.9 & 4.8 & 16.5 & 3.8
 & \multicolumn{2}{c}{\multirow{-7}{*}{}} \\ \hline
 & $\hat V_\text{Naïve}$  & 1.1 & 0.8 & 0.4 & 0.3 & 0.2 & 0.2 & 0.6 & 2.3 & 0.5 \\
 & $\hat V_\text{PARAM1}$ & 34.1 &  19.9 & 49.7 & 0.4 & 54.3 & 46.7 & 0.6 & 7.2 & 0.4  \\ 
 & $\hat V_\text{PARAM1(M)}$ & 0.3 & 0.3 & 0.3 & 0.3 & 0.3 & 0.3 & 0.3 & 0.3 & 0.3 \\ 
 &  $\hat V_\text{PARAM2}$ & 2.2 & 1.2 & 1.0 & 0.8 & 0.7 & 0.8 & 0.7 & 4.0 & 1.1\\ 
 &  $\hat V_\text{PARAM2(M)}$ & 1.9	& 10.0 & 7.8 &  0.6 &  0.5 & 4.8 & 8.6 & 9.3 & 14.0\\ 
 &  $\hat V_\text{NONPARAM}$ & 1.7 & 1.3 & 0.7 & 0.4 & 0.6 & 0.4 & 0.8 & 3.5 & 0.8\\ 
\multirow{-7}{*}{517210} & $\hat V_\text{NONPARAM(M)}$ & 2.0 & 1.3 & 0.8 &	0.7 & 0.6 & 0.7 & 0.7 & 4.0 & 1.0 \\ \hline
 &  $\hat V_\text{Naïve}$  & 3.4 & 4.0 & 5.9 & 6.8 & 6.5 & 23.9 & 3.0 & \multicolumn{2}{c}{} \\ 
& $\hat V_\text{PARAM1}$ & 3.5 & 5.5 & 18.6 & 7.9 & 9.1 & 29.9	& 4.7 & \multicolumn{2}{c}{} \\ 
 & $\hat V_\text{PARAM1(M)}$ & 2.9 & 2.8 & 3.2 & 2.9 & 2.9 & 3.0 & 2.9 & \multicolumn{2}{c}{} \\ 
 &  $\hat V_\text{PARAM2}$ & 3.2 & 4.8 & 8.7 & 7.3 & 8.4 &  23.4 & 4.2 & \multicolumn{2}{c}{} \\ 
 &  $\hat V_\text{PARAM2(M)}$ & 3.6 & 44.1 & 23.3 & 8.0 & 60.3 & 1035.5 & 7.6 & \multicolumn{2}{c}{} \\ 
 &  $\hat V_\text{NONPARAM}$ & 3.4 & 4.3 & 6.8 & 7.3 & 7.0 & 22.3 & 4.0 & \multicolumn{2}{c}{} \\ 
 \multirow{-7}{*}{541211} & $\hat V_\text{NONPARAM(M)}$ & 3.4 & 4.6 & 7.8 & 7.3 & 8.1 & 21.9 & 3.4 
 & \multicolumn{2}{c}{\multirow{-7}{*}{}} \\ \hline
 &  $\hat V_\text{Naïve}$  & 4.3 & 2.6 & 5.4 & 8.5 & 3.7 & 11.4 & 5.7 & 9.5 & 3.6 \\ 
  & $\hat V_\text{PARAM1}$ & 4.1 & 3.8 & 11.8 & 16.0 & 7.3 & 13.4 & 7.1 & 12.4 & 4.5 \\ 
 & $\hat V_\text{PARAM1(M)}$ & 3.0 & 3.3 & 3.2 & 3.2 & 3.3 & 3.0 & 3.2 & 3.2 & 3.1 \\ 
 &  $\hat V_\text{PARAM2}$ & 4.3 & 7.0 & 11.5 & 10.0 & 8.0 & 13.4 & 5.8 & 10.1 & 7.6 \\ 
 &  $\hat V_\text{PARAM2(M)}$ & 19.1 & 7.1 & 12.2 & 62.3 & 10.0 & 106.6 & 60.0 & 121.3 & 15.7 \\ 
 &  $\hat V_\text{NONPARAM}$ & 4.3 & 6.6 & 10.1 & 9.2 &	8.4 &	13.5 & 5.9 & 10.9 & 6.8 \\ 
\multirow{-7}{*}{621410} & $\hat V_\text{NONPARAM(M)}$ & 3.7 &	4.2 & 10.2 & 9.5 & 5.1 & 14.1 & 6.4 & 9.8 & 5.1\\ \hline
  &  $\hat V_\text{Naïve}$  & 1.4 & 0.8 & 0.5 & 1.0 & 1.3 & 
  4.2	& 1.3 & \multicolumn{2}{c}{} \\ 
  & $\hat V_\text{PARAM1}$ & 93.9 & 15.8 & 24.2 & 15.5 & 12.8 &	167.7 & 11.9 & \multicolumn{2}{c}{} \\ 
 & $\hat V_\text{PARAM1(M)}$ & 0.9 & 0.9 & 1.0 & 0.8 & 0.9  & 0.8 & 0.9 & \multicolumn{2}{c}{} \\ 
 &  $\hat V_\text{PARAM2}$ & 3.3 & 4.6 & 9.2 & 17.5 & 2.0 & 11.5 & 2.9 & \multicolumn{2}{c}{} \\ 
 &  $\hat V_\text{PARAM2(M)}$ & 1.4	& 3.1 &	7.1 & 219.3 & 3.5 & 30.3 & 1.8 & \multicolumn{2}{c}{} \\ 
 &  $\hat V_\text{NONPARAM}$ & 1.6 & 1.3 & 0.5 & 7.2 & 2.1 & 8.0 & 1.9  & \multicolumn{2}{c}{} \\ 
 \multirow{-7}{*}{713110} & $\hat V_\text{NONPARAM(M)}$ & 3.5 & 4.5 &	9.2 & 17.5 & 1.9 & 11.3 & 2.9  & \multicolumn{2}{c}{\multirow{-7}{*}{}} \\ \hline
\end{longtable}

\normalsize
As one might expect, given the results in Table \ref{tab:emp_ratio},  the c.v.'s obtained using $\hat V_\text{PARAM1}$ or $\hat V_\text{PARAM2(M)}$ estimates are much larger than those obtained with any other considered variance estimator. Recall that the PARAM1 model utlizes an \emph{industry level} ratio estimator, likely inappropriate. Table \ref{tab:emp_ratio} provides further indications that the modeled residuals in ($\hat V_\text{PARAM1(M)}$) do not improve this ratio estimator's performance, as the associated c.v.'s tend to be smaller than their \text{naïve} variance counterparts.  In contrast, the c.v.'s obtained with the $\hat V_\text{PARAM2}$, $\hat V_\text{NONPARAM}$, and $\hat V_\text{NONPARAM(M)}$ estimators are generally similar, with a few visible exceptions. These large differences could be due to model misspecifications, but could also be confounded with small sample size effects for many of the detail items. 

Despite the similarity of their c.v.'s, the variance estimates of corresponding items obtained with $\hat V_\text{PARAM2}$, $\hat V_\text{NONPARAM}$, and $\hat V_\text{NONPARAM(M)}$ are quite different. This affects confidence interval width, and expected coverage by extension. Without a “gold standard” against  which  to  measure  these variances, however, we have no viable recommendation.  Consequently,  we  conducted the  Monte  Carlo simulation  study described in Section \ref{sec:simulation}. 

\section{Simulation Study}\label{sec:simulation}

To evaluate the finite-sample performance of the proposed method over repeated samples, we conducted a simulation study.  The simulation varies in four factors: parametric distribution of the size (auxiliary) variable $x_i$ , the size of the finite population ($N$), relationship of auxiliary variable and detail items ($x_{i}$ and $y_{i}$), and response propensity. The data generation is largely patterned after the realistic procedures described in \citet*{andridge2020finding}, with each separate process outlined below.

\subsection{Create and Stratify the Finite Population}

We generated \emph{three} different sets of $B=2000$ finite populations of size $N$ by drawing the size (auxiliary) variable $x_{i}$ from

\textbf{Population Scenario 1:} $x_i \sim 100,000*U(0,1)$

\textbf{Population Scenario 2:} $x_i \sim Lognormal(4.1, 0.66)$

\textbf{Population Scenario 3:} $x_i \sim Lognormal(12, 1.72)$

The first population scenario ensures the compact and convex support requirements of Assumption \ref{ass:regular}. Thus, these data represent ideal conditions for the proposed NNRI variance estimators. However, business data population such as the SAS industry populations discussed in Section \ref{sec:application} are generally positively skewed. Consequently, we consider two lognormal distributions. The second population scenario exhibits mild deviations from the required compact support requirement, while respecting the convex support requirement (similar to the 517210, 541210, and 713110 industries' total expenses distributions discussed in Section \ref{sec:application}). The third population scenario creates finite populations that do not exhibit the compact support requirement of Assumption \ref{ass:regular}, but resemble 221122 and 621410 industries' total expenses distributions presented in Section \ref{sec:application}. Thus, the two lognormal population scenarios provide insights into the empirical results while testing the robustness of the proposed variance estimation approach.

Each finite population is stratified using the strata boundaries provided in Table \ref{tab:strata}. 

\begin{table}
\caption{\label{tab:strata}{Strata boundaries for the Simulation Study (Section \ref{sec:simulation})}}
\begin{centering}
\begin{tabular}{llll}
\hline 
Stratum $S$ & Population Scenario 1 & Population Scenario 2 & Population 3\tabularnewline
\hline 
1 & $<25,000$  &{$<55$}  & $<40,000$\tabularnewline
2  & $25,000\leq X < 50,000$ & {$55\leq X <85$} & $40,000\leq X < 150,000$ \tabularnewline
3  & $50,000\leq X < 75,000$& {$85\leq X <150$} & $150,000\leq X < 500,000$ \tabularnewline
4 & $\geq 75,000$ &{$ \geq 150$} &$\geq 500,000$ \\
\bottomrule
\end{tabular}
\par\end{centering}
\bigskip{}
\end{table}
 
\subsection{Generate Sets of Detail Items in Stratified Finite Populations}

We used a two-step process to generate the sets of detail items values $y_{i}=(y_{i1},y_{i2},y_{i3},y_{i4},y_{i5})^{\intercal}$ associated with each unit, capturing important data features observed in several of the U.S. Census Bureau's economic programs. Specifically, the $y_{i}$ for all units have non-zero values assigned to $(y_{i1},y_{i2})$ but may have zero values in $(y_{i3},y_{i4},y_{i5})$.  

To justify the use of the NNRI procedure, the \emph{number of non-zero detail items} is directly related to unit size. Within each stratum $S_i$, we generate $C_{i}$ (the number of non-zero detail items) for unit $i$ from a discrete distribution of \{2,3,4,5\} with selection probability $P({C_i=c} \mid x_{i})$ = $p(x_i)$ where $p(x)=(\pi_{1},\pi_{2},\pi_{3},\pi_{4},\pi_{5})$ are given by
 \[
(\pi_{1},\pi_{2},\pi_{3},\pi_{4},\pi_{5})=\begin{cases}
(0,.91,.03,.03,.03) & \text{in Stratum 1}\\
(0,.50,.40,.05,.05) & \text{in Stratum 2}\\
(0,.20,.20,.30,.30) & \text{in Stratum 3}\\
(0,.05,.15,.40,.40) & \text{in Stratum 4}
\end{cases}.
\]

\begin{figure}[htbp]
    \centering
    \includegraphics[width=.8\linewidth]{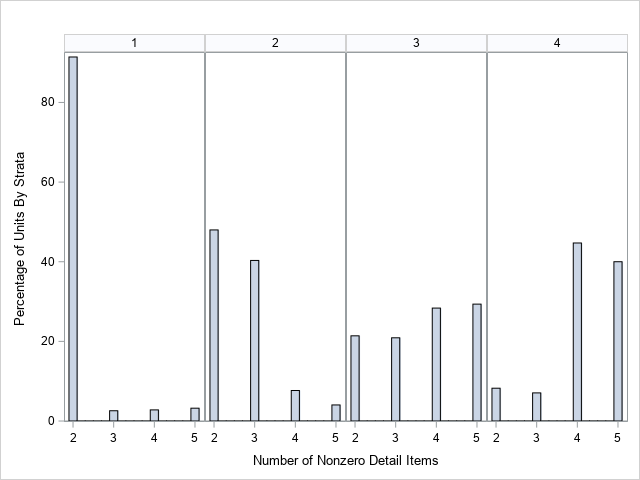}
    \caption{Distribution of Realized Nonzero Detail Items $c$ in a Single Simulated Finite Population. Histograms are computed within sampling strata.  Strata are numbered in increasing order, with Stratum 1 containing the smallest units.}
    \label{Cs:sim}
\end{figure}

Figure \ref{Cs:sim} presents histograms of the realized values of the number of nonzero details ($c$) from a single simulated population. As the unit size ($x$) increases, the number of nonzero detail items reported by each unit tends to likewise increase: for example, in the smallest unit size stratum ($1$), the majority of units provide two nonzero values, whereas in the largest unit size stratum ($4$), the majority of units provide four or five nonzero values.

Conditioning on the assigned $c_{i}$, we draw the $R_{i}$ for each unit $i$ from a multinomial distribution.
\[
R_{i}\mid(x_{i},c_{i})\sim\text{Multinomial}\{x_{i},(p_{1},p_{2},p_{3},p_{4},p_{5})\},
\]
with probabilities
\[
(p_{1},p_{2},p_{3},p_{4},p_{5})=\begin{cases}
(.60,.40,.00,.00,.00) & \text{if }c=2\\
(.60,.30,.10,.00,.00) & \text{if }c=3\\
(.60,.25,.10,.05,.00) & \text{if }c=4\\
(.60,.20,.10,.05,.05) & \text{if }c=5
\end{cases}.
\]

\begin{figure}[htbp]
    \centering
    \includegraphics[width=.8\linewidth]{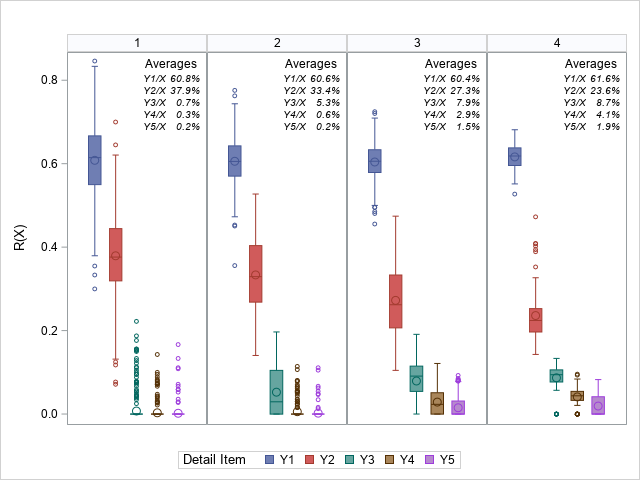}
    \caption{Distribution of Detail Item Ratios $R(x)$ by Sampling Strata in a Single Simulated Finite Population. Average values are computed within sampling strata.}
    \label{RatioSim}
\end{figure}

By design, $p_{1}$ = 0.60 for all units regardless its class. This detail item therefore represents about $60\%$ of each unit's total ($ x_{i}$). Figure \ref{RatioSim} illustrates the subtle change in multinomial distributions as unit size (sampling strata) increases. The largest proportion of the total is always reported in item $Y1$, with item $Y2$ following. The remaining three items are more rarely reported, with the probability of a reported nonzero value being strongly related to unit size. This mimics patterns that were observed by \cite{andridge2020finding} in several economic census datasets.

Lastly, using these $R_{i}$, we compute $y_{i}=(y_{i1},y_{i2},y_{i3},y_{i4},y_{i5}) = x_{i}(R_{i1},R_{i2},R_{i3},R_{i4},R_{i5}).$ This ensures that $x_{i}=\sum_{t=1}^{5}y_{it}$.

\subsection{Select Stratified SRS-WOR Samples}

We select a single stratified SRS-WOR sample from each finite population. Table \ref{tab:The-sample-allocation} provides the finite population size ($N$), the {average} stratum sizes $N_h$, the sampling fractions $f_h$, and {average} sample sizes $n_h$. Notice that the overall sampling fraction ($f$) is $302/1000$ in the $N=1000$ populations and is $152/500$ in the $N=500$ populations and are therefore both non-negligible. As typical of business surveys, the largest units are grouped into a certainty stratum, and a high proportion of the medium-sized units (grouped into strata 2 and 3) are sampled at a high (non-negligible) rates, requiring inclusion of the finite-population correction in $\hat V^m$.

\begin{table}
\caption{\label{tab:The-sample-allocation}{Averaged Stratum size and sample allocations. Sampling fractions are fixed in all simulations}}
\begin{centering}
\begin{tabular}{lcccc}
\hline 
Stratum $S$ & 1 & 2 & 3 & 4\tabularnewline
\hline 
$N=1000$\\ 
$N_{h}$ & 442 &255  &  219& 83\tabularnewline
$f_{h}$ & $1/10$ & $1/4$ & $1/2$ & $1$\tabularnewline
$n_{h}$ & 45 & 64 & 110 &83 \tabularnewline
\hline 
\end{tabular}
\begin{tabular}{lcccc}
\hline 
Stratum $S$ & 1 & 2 & 3 & 4\tabularnewline
\hline 
$N=500$\\
$N_{h}$ & 221 & 128 &110  &42\tabularnewline
$f_{h}$ & $1/10$ & $1/4$ & $1/2$ & $1$\tabularnewline
$n_{h}$ & 23 & 32 & 55 & 42\tabularnewline
\hline 
\end{tabular}
\par\end{centering}
\bigskip{}
\end{table}

\subsection{Induce Nonresponse and Impute}

 Finally, within each sample, we generate missingness indicators for each unit $i$ as $\delta_{i}\sim\textit{Bernoulli}(\pi)$ under different response mechanisms: uniform (missing completely at random) with $\pi$ = $75\%$ and $50\%$ (MCAR); uniform response within strata (missing at random) with smaller units being more likely to respond (Negative MAR); and uniform response within strata with larger units being more likely to respond (Positive MAR). Table \ref{tab:MAR} presents the strata response propensities for the negative and positive MAR response mechanisms. Notice that between-stratum differences in response propensities quite large; these discrepancies are exaggerated for illustration. The MCAR response propensities resemble the observed patterns for industry 713110 and to a lesser extent, for industries 541211 and 621410 presented in Section \ref{sec:application}, although the latter two industries could equally be categorized with negative MAR response patterns. The negative MAR response mechanism mimics the observed pattern for industry 221122, whereas the positive MAR response mechanism mimics the observed pattern for industry 517210. 
 
\begin{table}
\caption{\label{tab:MAR}{Response propensities for the negative and positive MAR mechanisms}}
\begin{centering}
\begin{tabular}{lcccc}
\hline 
Negative MAR\\ 
\hline 
Stratum $S$ & 1 & 2 & 3 & 4\tabularnewline
$\pi_h$ & 0.85 & 0.65  &  0.45 & 0.25 \tabularnewline
\hline 
\end{tabular}
\begin{tabular}{lcccc}
\hline 
Positive MAR\\
\hline 
Stratum $S$ & 1 & 2 & 3 & 4\tabularnewline
$\pi_h$ & 0.25 & 0.45 & 0.65  & 0.85\tabularnewline
\hline 
\end{tabular}
\par\end{centering}
\bigskip{}
\end{table}
 
 Imputation and estimation are performed separately within each strata, with $x_i$ as the matching variable for the NNRI of ${{y}_{i}}=(y_{i1},\cdots,y_{iT})$ as outlined in Section \ref{subsec:NNRI}.
 
\subsection{Simulation Results}\label{subsec:simresults}

To evaluate the performance of the proposed NNRI variance estimators over repeated samples selected from different population distributions (scenarios), finite population sizes, and response mechanisms, we compute the relative bias of each variance estimator 
$\widehat V_{yp}$ and the coverage rates of the approximate $95\%$ confidence intervals constructed with $\widehat T_y$ and $\widehat V_{yp}$ for variance estimation method $p$. 

Table \ref{tab:RelBiases} reports the relative biases using \emph{directly obtained residuals} for all five detail items for each population scenario and response mechanism for the $N=1000$ populations. Results for the $N=500$ populations are similar and are consequently not presented here, but are available upon request to the authors. The relative bias of each variance estimator is computed as $\text{RB}(\widehat V^{(b)}_{yp}$) 
$=[({\sum_{b=1}^B{\widehat V^{(b)}_{yp}}/{B}})/{ V_{yp}}] - 1$ where $\widehat V^{(b)}_{yp}$ is the estimate from the $b^{th}$ sample ($B=2000$) and $V_{yp}$ is the Monte Carlo empirical (true) variance. Recall that  $\sum_{i=2}^5 Y_i$ represents a \emph{maximum} of 40\% of the corresponding total, $\sum_{i=3}^5 Y_i$ represents a \emph{maximum} of 20\% of the corresponding total, and $\sum_{i=4}^5 Y_i$ representS 5\% orless of the corresponding total. Consequently, the analysis of relative biases of the variance estimates will be confounded with small sample size effects for each detail item except $Y_1$, with small unit differences in totals potentially representing large percentage differences. The main results of Table \ref{tab:RelBiases} can be summarized as follows: 

\begin{table}
  \caption{
  \label{tab:RelBiases}Relative biases of variance estimates using \emph{directly-obtained residuals} for all detail items by population scenario and response mechanism computed from $2,000$ independent stratified SRS-WOR samples from the $N=1000$ populations. \emph{Negative} relative biases are in parenthesis. Population Scenario 1 = Uniform; Population Scenario 2 = Lognormal ($\mu=4.1$, $\sigma = 0.66$); Population Scenario 3 = Lognormal ($\mu=12.0$, $\sigma = 1.7$)}
  \centering
  \tiny
  \resizebox{\textwidth}{!}{
    \begin{tabular}{lllllll|lllll}
    \toprule
    \multicolumn{1}{l}{\multirow{2}[4]{*}{\makecell{Population\\Scenario}}} & \multicolumn{1}{l}{\multirow{2}[4]{*}{ \makecell{Method\\$(\widehat{V}_{yp})$}}} & \multicolumn{5}{c}{MCAR $(\pi= 0.75)$} & \multicolumn{5}{c}{MCAR $(\pi= 0.50)$} \\
\cmidrule{3-12}          &       & ${Y_1}$ &  ${Y_2}$  &  ${Y_3}$  &  ${Y_4}$  &  ${Y_5}$   & ${Y_1}$ &  ${Y_2}$  &  ${Y_3}$  &  ${Y_4}$  &  ${Y_5}$ \\
    \midrule
    \multicolumn{1}{l}{\multirow{4}[2]{*}{\makecell{Population\\Scenario 1} }} &  NAIVE  & (0.00)  & (0.77) & (0.89) & (0.93) & (0.95) & (0.01) & (0.54) & (0.75) & (0.82) & (0.85) \\
          &  PARAM1  & (0.00)  & 0.45  & 0.47  & 0.56  & 0.12  & (0.00)  & 0.31  & 0.51  & 0.64  & 0.23  \\
          &  PARAM2  & (0.00)  & (0.04) & (0.07) & (0.07) & (0.05) & (0.00)  & (0.02) & (0.05) & (0.03) & 0.03  \\
          &  NONPARAM  & 0.01  & (0.02) & 0.02  & (0.01) & (0.04) & 0.00  & (0.05) & 0.04  & 0.03  & 0.05  \\
    \midrule
    \multicolumn{1}{l}{\multirow{4}[2]{*}{\makecell{Population\\Scenario 2}}} &  NAIVE  & (0.13) & (0.36) & (0.49) & (0.57) & (0.62) & (0.29) & (0.57) & (0.73) & (0.78) & (0.82) \\
          &  PARAM1  & (0.00)  & 0.14  & 0.70  & 0.54  & 0.18  & 0.02  & 0.33  & 0.67  & 0.58  & 0.18  \\
          &  PARAM2  & (0.01) & (0.07) & (0.02) & (0.04) & (0.03) & 0.00  & (0.01) & (0.02) & (0.00)  & (0.02) \\
          &  NONPARAM  & 0.02  & (0.12) & 0.11  & 0.00  & (0.04) & 0.03  & (0.03) & 0.09  & 0.01  & (0.07) \\
    \midrule
    \multicolumn{1}{l}{\multirow{4}[2]{*}{\makecell{Population\\Scenario 3}}} &  NAIVE  & 0.04  & (0.96) & (0.96) & (0.99) & (1.00) & 0.04  & (0.98) & (0.98) & (1.00) & (1.00) \\
          &  PARAM1  & 0.04  & (0.22) & (0.08) & (0.35) & (0.28) & 0.04  & (0.08) & (0.02) & (0.15) & (0.20) \\
          &  PARAM2  & 0.04  & (0.28) & (0.17) & (0.39) & (0.30) & 0.04  & (0.14) & (0.10) & (0.19) & (0.21) \\
          &  NONPARAM  & 0.09  & (0.73) & (0.57) & (0.78) & (0.82) & 0.21  & (0.76) & (0.69) & (0.77) & (0.81) \\
    \midrule
          &         & \multicolumn{5}{c}{Negative MAR}      & \multicolumn{5}{c}{Positive MAR} \\
\cmidrule{3-12}           &       & ${Y_1}$ &  ${Y_2}$  &  ${Y_3}$  &  ${Y_4}$  &  ${Y_5}$   & ${Y_1}$ &  ${Y_2}$  &  ${Y_3}$  &  ${Y_4}$  &  ${Y_5}$ \\
    \midrule
    \multicolumn{1}{l}{\multirow{4}[2]{*}{\makecell{Population\\Scenario 1}}} &  NAIVE  & (0.01) & (0.60) & (0.75) & (0.88) & (0.92) & (0.00)  & (0.49) & (0.72) & (0.74) & (0.75) \\
          &  PARAM1  & (0.01) & 0.45  & 0.64  & 0.57  & 0.11  & (0.00)  & 0.68  & 1.08  & 1.49  & 0.46  \\
          &  PARAM2  & (0.01) & 0.02  & (0.00)  & (0.02) & (0.03) & (0.00)  & (0.06) & (0.02) & (0.07) & (0.06) \\
          &  NONPARAM  & 0.00  & (0.02) & 0.06  & (0.00)  & (0.04) & 0.00  & (0.07) & 0.16  & 0.06  & 0.02  \\
    \midrule
    \multicolumn{1}{l}{\multirow{4}[2]{*}{\makecell{Population\\Scenario 2} }} &  NAIVE  & (0.22) & (0.52) & (0.72) & (0.83) & (0.88) & (0.44) & (0.68) & (0.77) & (0.77) & (0.75) \\
          &  PARAM1  & (0.01) & 0.53  & 0.81  & 0.67  & 0.16  & (0.01) & 0.59  & 1.32  & 1.23  & 0.56  \\
          &  PARAM2  & (0.02) & (0.01) & (0.04) & (0.03) & (0.05) & (0.05) & (0.11) & (0.09) & (0.09) & (0.02) \\
          &  NONPARAM  & (0.01) & (0.10) & (0.03) & (0.15) & (0.26) & (0.01) & (0.04) & 0.32  & 0.06  & 0.07  \\
    \midrule
    \multicolumn{1}{l}{\multirow{4}[2]{*}{\makecell{Population\\Scenario 3}}} &  NAIVE  & 0.04  & (0.99) & (0.99) & (1.00) & (1.00) & 0.04  & (0.94) & (0.93) & (0.99) & (0.99) \\
          &  PARAM1  & 0.04  & (0.11) & (0.04) & (0.14) & (0.24) & 0.04  & (0.26) & (0.07) & (0.40) & (0.30) \\
          &  PARAM2  & 0.04  & (0.21) & (0.17) & (0.21) & (0.28) & 0.04  & (0.33) & (0.19) & (0.45) & (0.33) \\
          &  NONPARAM  & 0.46  & (0.86) & (0.83) & (0.84) & (0.88) & 0.08  & (0.69) & (0.47) & (0.76) & (0.82) \\
    \bottomrule
    \end{tabular}%
    }
\end{table}%

\begin{itemize}
\item The \text{naïve} variance estimator severely underestimates the true variance of detail items $Y_2$, $Y_3$, $Y_4$, and $Y_5$ for all populations and response mechanisms;
\item In Population Scenarios 1 and 2, the PARAM1 variance estimator \emph{overestimates} the true variance of all detail items except for $Y_1$, essentially confirming the conjecture of overestimation posited Section \ref{sec:application} for industries 517210, 541210, and 713110. This variance estimator generally \emph{underestimates} the true variance of the same detail items in Population Scenario 3, mimicking the empirical results for industries 221122 and 631400.
\item In Population Scenarios 1 and 2, the PARAM2 variance estimates are nearly unbiased, regardless of response mechanism. However, in Population Scenario 3, the PARAM2 variance estimator tends to underestimate the variance of all detail items except for $Y_1$, with the degree of underestimation being more severe than their PARAM1 variance estimate counterparts.
\item In Population Scenarios 1 and 2, the NONPARAM variance have inconsistent relative bias performance, although generally improved over the corresponding \text{naïve} variance estimates. This is not true in Population Scenario 3, where the variances estimates for all detail items except $Y_1$ are severely underestimated, and the variance estimates for $Y_1$ are overestimates, with the level of overestimation related to the response mechanism.
\end{itemize}

The simulation conditions in Population 2 with a MCAR or negative MAR response mechanism resemble the 541210 and 713110 industries' conditions; Table \ref{tab:RelBiases} provides evidence that the $\hat V_{PARAM2}$ are likely the most accurate estimates in this situation, even for the rarely-reported detail items. The simulation conditions in Population 2 with a positive MAR response mechanism resemble the industry 517210 conditions; the simulation results are close to the same for the $\hat V_{PARAM2}$ and the $\hat V_{NONPARAM}$, without a clear-cut favorite. The simulation conditions in Population 3 with a negative MAR response mechanism resemble the 221122 and 621400 industries' data; the relative biases for the MCAR response mechanism are similar for $\hat V_{PARAM1}$ and $\hat V_{PARAM2}$, but the $\hat V_{PARAM2}$ are less biased under the negative MAR response mechanism (the $\hat V_{NONPARAM}$ are severe underestimates for all detail items).

Table \ref{tab:RelBiasesM} reports the relative biases using the \emph{modeled residuals} (see Section \ref{subsec:Ve}) for all five detail items for each population scenario and response mechanism for the $N=1000$ populations. With the ``high'' uniform response rate (i.e., MCAR with $\pi = 0.75$), as well as the negative and positive MAR response mechanism, modeling the residuals for ${\sigma}_e^{2}(x_{i})$ generally reduces the relative bias of the PARAM2 variance estimates. Notice that the PARAM2(M) variance estimates are nearly unbiased in Population Scenario 3 for all detail items, except the lowest uniform response rate mechanism (MCAR with $\pi=0.50)$; in this situation, the bias effects are likely overstated for the most rarely reported detail items (e.g, $Y_4$ and $Y_5$). This results contrast with those presented in the empirical case study in Section \ref{sec:application}, and we suspect this could be an artifact of the data simulation process. The other approaches (PARAM1(M) and NONPARAM(M)) do not yield consistent improvements over their counterparts obtained with directly-obtained residuals. Given that the $2^{nd}$ parametric method of estimating $m_i$ (PARAM2) generally yields accurate variance estimates and the associated procedures for obtaining $\widehat \sigma^2_e$ appear tractable, we dropped the alternative parametric approach (PARAM1) from further consideration.

\begin{table}
  \caption{
  \label{tab:RelBiasesM}Relative biases of variance estimates using \emph{modeled residuals} for all detail items by population scenario and response mechanism computed from $2,000$ independent stratified SRS-WOR samples from the $N=1000$ populations. \emph{Negative} relative biases are in parenthesis. Population Scenario 1 = Uniform; Population Scenario 2 = Lognormal ($\mu=4.1$, $\sigma = 0.66$); Population Scenario 3 = Lognormal ($\mu=12.0$, $\sigma = 1.7$)}
  \centering
  \tiny
  \resizebox{\textwidth}{!}{
    \begin{tabular}{lllllll|lllll}
    \toprule
    \multicolumn{1}{l}{\multirow{2}[4]{*}{\makecell{Population\\Scenario}}} & \multicolumn{1}{l}{\multirow{2}[4]{*}{  \makecell{Method\\($\widehat{V}_{yp}$)}}} & \multicolumn{5}{c}{MCAR $(\pi= 0.75)$} & \multicolumn{5}{c}{MCAR $(\pi= 0.50)$} \\
\cmidrule{3-12}           &       & ${Y_1}$ &  ${Y_2}$  &  ${Y_3}$  &  ${Y_4}$  &  ${Y_5}$   & ${Y_1}$ &  ${Y_2}$  &  ${Y_3}$  &  ${Y_4}$  &  ${Y_5}$ \\
    \midrule
    \multicolumn{1}{l}{\multirow{3}[2]{*}{\makecell{Population\\Scenario 1} }} &  PARAM1(M)  & (0.00)  & (0.90) & (0.98) & (0.99) & (1.00) & 0.00  & (0.81) & (0.95) & (0.97) & (0.99) \\
          &  PARAM2(M)  & (0.00)  & (0.04) & (0.06) & (0.06) & (0.05) & (0.00)  & (0.01) & (0.04) & (0.02) & 0.04  \\
          &  NONPARAM(M)  & 0.01  & 2.42  & 16.03  & 0.14  & 0.06  & 0.00  & 0.07  & 0.48  & 0.18  & 0.16  \\
    \midrule
    \multicolumn{1}{l}{\multirow{3}[2]{*}{\makecell{Population\\Scenario 2}}} &  PARAM1(M)  & 0.00  & (0.47) & (0.46) & (0.23) & (0.47) & 0.02  & (0.47) & (0.50) & (0.30) & (0.55) \\
          &  PARAM2(M)  & (0.01) & (0.06) & (0.01) & (0.02) & (0.02) & 0.00  & (0.01) & (0.02) & 0.00  & (0.01) \\
          &  NONPARAM(M)  & 0.03  & (0.08) & 0.42  & 0.15  & 0.16  & 0.05  & 0.04  & 0.41  & 0.16  & 0.12  \\
    \midrule
    \multicolumn{1}{l}{\multirow{3}[2]{*}{\makecell{Population\\Scenario 3}}} &  PARAM1(M)  & 0.04  & (0.98) & (0.98) & (1.00) & (1.00) & 0.04  & (0.99) & (0.99) & (1.00) & (1.00) \\
          &  PARAM2(M)  & 0.04  & 0.10  & 0.34  & 0.00  & 0.05  & 0.04  & 0.22  & 0.29  & 0.22  & 0.10  \\
          &  NONPARAM(M)  & 0.20  & (0.40) & 0.26  & (0.51) & (0.65) & 0.41  & (0.55) & (0.22) & (0.57) & (0.69) \\
    \midrule
          &         & \multicolumn{5}{c}{Negative MAR}      & \multicolumn{5}{c}{Positive MAR} \\
\cmidrule{3-12}          &       & ${Y_1}$ &  ${Y_2}$  &  ${Y_3}$  &  ${Y_4}$  &  ${Y_5}$   & ${Y_1}$ &  ${Y_2}$  &  ${Y_3}$  &  ${Y_4}$  &  ${Y_5}$ \\
    \midrule
    \multicolumn{1}{l}{\multirow{3}[2]{*}{\makecell{Population\\Scenario 1} }} &  PARAM1(M)  & (0.00)  & (0.81) & (0.96) & (0.99) & (1.00) & (0.00)  & (0.81) & (0.93) & (0.95) & (0.98) \\
          &  PARAM2(M)  & (0.00)  & 0.01  & (0.00)  & (0.02) & (0.02) & (0.00)  & (0.05) & (0.01) & (0.04) & (0.02) \\
          &  NONPARAM(M)  & 0.02  & 0.00  & 0.15  & 0.04  & (0.01) & 0.04  & 0.54  & 10.88  & 0.40  & 0.18  \\
    \midrule
    \multicolumn{1}{l}{\multirow{3}[2]{*}{\makecell{Population\\Scenario 2} }} &  PARAM1(M)  & (0.01) & (0.49) & (0.68) & (0.71) & (0.84) & 0.01  & (0.46) & (0.24) & 0.48  & 0.25  \\
          &  PARAM2(M)  & (0.02) & (0.00)  & (0.03) & (0.02) & (0.04) & (0.05) & (0.09) & (0.08) & (0.06) & (0.00)  \\
          &  NONPARAM(M)  & 0.00  & (0.07) & 0.06  & (0.09) & (0.19) & 0.06  & 0.17  & 1.25  & 0.30  & 0.43  \\
    \midrule
    \multicolumn{1}{l}{\multirow{3}[2]{*}{\makecell{Population\\Scenario 3}}} &  PARAM1(M)  & 0.04  & (1.00) & (1.00) & (1.00) & (1.00) & 0.04  & (0.97) & (0.97) & (1.00) & (1.00) \\
          &  PARAM2(M)  & 0.04  & (0.05) & (0.02) & (0.02) & (0.07) & 0.04  & (0.03) & 0.17  & (0.10) & (0.00)  \\
          &  NONPARAM(M)  & 0.55  & (0.82) & (0.77) & (0.80) & (0.85) & 0.72  & 8.17  & 16.58  & 5.78  & 0.81  \\
    \bottomrule
    \end{tabular}%
    }
\end{table}%

\begin{figure}[htbp]
    \centering
    \includegraphics[width=.8\linewidth]{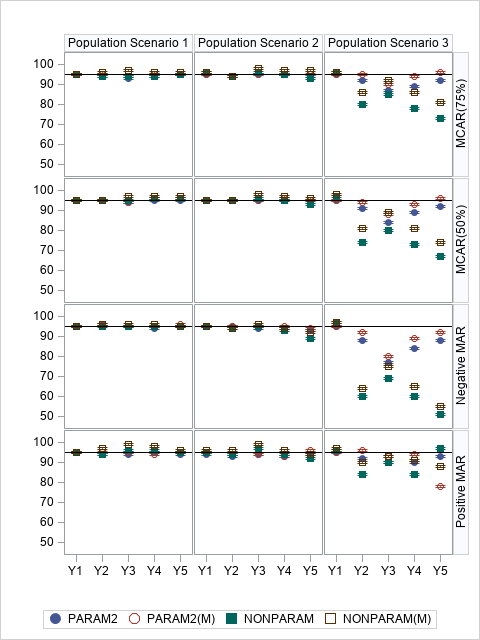}
    \caption{Coverage rates with Monte Carlo confidence bounds(in percentages) by population scenario and response mechanism using directly-obtained and modeled residuals with the PARAM2 and NONPARAM methods for the $N=1000$ populations. Nominal coverage indicated by horizontal asymptote.}
    \label{Cs:sim1}
\end{figure}

Regardless of population scenario, population size, and considered response mechanism, all NNRI \emph{totals} were unbiased across the repeated samples. Again, we conjecture that this is an artifact of the simulation design, which ensures appropriate conditions for the distance function used to select the nearest neighbor for imputation. Nevertheless, coverage rates provide a measure of the practical impact of the bias of the considered variance estimators, given the unbiased estimates. Figure \ref{Cs:sim1} presents the coverage rates using confidence intervals constructed from the PARAM2, PARAM2(M), NONPARAM, and NONPARAM(M) variance estimates.

Figure \ref{Cs:sim1} provides evidence of
\begin{itemize}
\item Nominal or slight undercoverage with the PARAM2 variance estimates regardless of response mechanism when Assumption \ref{ass:regular} is fully met (Population Scenario 1) or is mildly violated (Population Scenario 2). Moderate undercoverage under strong violations of Assumption \ref{ass:regular}, regardless of response mechanism. However, the undercoverage is nearly abated with the PARAM2(M) variances, except for the most infrequently reported item ($Y_5$) with the positive MAR response mechanism. In this case, the modeled residuals are inadequate, probably because small units are less likely to respond under this response mechanism and $Y_5$ is (by design) rarely reported by smaller units. 
\item Inconsistent but rarely nominal coverage for the NONPARAM variance estimates for all response mechanisms in Population Scenarios 1 and 2 and consistent overcoverage with the NONPARAM(M) variance estimates in the same scenarios. Severe undercoverage with NONPARAM variance estimates for all response mechanisms in Population Scenario 3, with no improvements offered by the NONPARAM(M) approach. 
\end{itemize}

Taken collectively, the simulation results support the generally poor performance of the PARAM1 and PARAM1(M) methods demonstrated in Section \ref{sec:application}. Ultimately, the results obtained with PARAM2 are promising, especially given that the data generation models were not congenial to the \emph{WLS} regression used to obtain $\widehat m_i$ and the data violations in the third data generation scenario (found in two of the five studied empirical distributions). Despite its poor performance in this simulation study, the nonparametric method remains appealing due to its flexbility. It is possible that the model fit and estimation might be improved with a different choice of basis functions, although we would not recommend utilizing either variation with the studied industries in Section \ref{sec:application}. That said, some caution should be exercised in over-generalizing these results, as the simulation utilizes a very specific multinomial distribution and a single sampling design. 

\section{Concluding remarks}\label{sec:Conclusion}
Nearest neighbor ratio imputation (NNRI) is a useful approach for imputing an \emph{entire set} of component detail items. 
Instead of directly imputing the set of detail item values ($y_i$) from the donor, NNRI imputes the proportions of donor ratios ($R_i$), which are in turn multiplied by the recipient's available total to derive imputed values for all items. This imputation method has several appealing properties, especially from a bias reduction perspective as discussed in Section 1. 

NNRI guarantees additivity, as the summed details always equal the associated total.  It accommodates subtle changes in unit level multinomial distributions that are associated with unit size. It yields realistic microdata, preserving multivariate relationships. In contrast to other frequently used imputation methods, the NNRI avoids inadvertently imposing possibly outdated historical patterns in the imputed data, as it is entirely restricted to current data \citep{andridge2020finding}. 

The numerous advantages of the NNRI procedure can be offset by the difficulty of obtaining a valid variance estimator. However, \citet{yang2019nearest} show that by  identifying the nearest donor using a single scalar with a suitable distance function, the NNRI estimator is asymptotically consistent. Following the frameworks of \citet{shao1999variance} and \citet{yang2019nearest}, we decompose our asymptotic variance into two parts and extend the variance estimation further to include the case of non-negligible sampling fractions %\st{through}
employing both parametric and nonparametric models to obtain smooth estimators for set of ratios. Selecting an appropriate model for the data set at hand is essential, as demonstrated by the empirical application and the simulation study. Model determination is not completely straightforward: in both the empirical application and our simulation, the frequently reported detail items tend to be strongly associated with the total, whereas the relationship between rarely-reported detail items and the total is less obvious. Nevertheless, our simulation studies provide fairly promising results in terms of bias and coverage, even with some model misspecification. In practice, one would expect that methodologists would develop and validate any implemented models after careful data analysis before implementing the proposed variance estimator. 

Although promising, the empirical and simulation study results collectively suggest several areas of future research. First, we limited ourselves to uniform and lognormally distributed size variables in our study and restricted the response mechanisms to tractable MCAR or MAR models.  Thus, the sensitivity of our variance estimator to more complex MAR models or even to  non-random response mechanism bears study. Second, if the proportion of recipients to donors is large, then NNRI may repeatedly use the same donor,  yielding insufficient variation within each imputation cell.  \citet{andridge2020finding} propose a modification of the NNRI method that addresses this issue in a multiple imputation framework; it would be useful to develop a single imputation analogue. 
Third, in practice, many auxiliary variables can be used to determine nearest neighbors, in which case, dimension reduction is necessary to mitigate matching discrepancies. Several techniques such as propensity score \citep{rosenbaum&rubin83a}, prognostic score \citep{hansen2008prognostic}, or their combination \citep{yang2020multiply} can be potentially adopted in the NNRI framework.  
Finally, this paper considers only population totals. However, extending the current framework to general parameter estimation is also feasible \citep{yang2020asymptotic}. Given that hot deck imputation is used to create realistic microdata (as well as macrodata), such extensions are especially compelling topics for our future research.
\section*{Acknowledgements}

We thank Carol Caldwell, William Davie Jr., Matthew Thompson, and Katrina Washington for their helpful comments on earlier versions of this manuscript, and Stephen Kaputa for his contributions to the empirical analysis. Yang is partially supported by the NSF grant DMS 1811245, IA grant 1R01AG06688, and NIEHS grant 1R01ES031651. Kim is partially supported by the NSF grant MMS 1733572.

\bibliography{NNRI}
\bibliographystyle{chicago}

\newpage

\begin{center}
\textbf{\large Supporting Information for ``Nearest neighbor ratio imputation with incomplete multi-nomial outcome in survey sampling'' by Gao et al.}{\Large\par}
\par\end{center}
\pagenumbering{arabic}
\global\long\def\theequation{S.\arabic{equation}}
\setcounter{equation}{0}
\global\long\def\thesection{S.\arabic{section}}
\setcounter{section}{0}
\global\long\def\thelemma{S.\arabic{lemma}}

\section{Proof of (\ref{eq:kappa}) and Lemma \ref{lem:asym_equal}}
To demonstrate (\ref{eq:kappa}), we write 
\begin{align*}
    \sum_{i\in\mathcal{S}}\delta_i w_i(1+\kappa_i)x_i&=
    \sum_{i\in\mathcal{S}}\delta_i w_i x_i+
    \sum_{i\in\mathcal{S}}\delta_i w_i\kappa_i x_i\\
    &=\sum_{i\in\mathcal{S}}\delta_i w_i x_i+\sum_{i\in \mathcal{S}}\delta_i w_ix_i \sum_{j\in\mathcal{S}}\frac{w_jx_j}{w_ix_i}(1-\delta_j)d_{ij}\\
    &=\sum_{i\in\mathcal{S}}\delta_i w_i x_i+\sum_{i\in \mathcal{S}} \sum_{j\in\mathcal{S}}\delta_i w_jx_j(1-\delta_j)d_{ij}\\
    &=\sum_{i\in\mathcal{S}}\delta_i w_i x_i+\sum_{j\in \mathcal{S}}w_jx_j(1-\delta_j) \sum_{i\in\mathcal{S}}\delta_i d_{ij}\\
    &=\sum_{i\in\mathcal{S}} w_i x_i,
\end{align*}
where the equivalence from the third line to the fourth line implies that $\sum_{i\in\mathcal{S}}\delta_i d_{ij}=1$, i.e., for every nonrespondent $j$, there always exist a donor $i\in\mathcal{S}$ s.t. $d_{ij}=1$. 
To demonstrate Lemma \ref{lem:asym_equal}, using algebra, we obtain 
\begin{align}
&\sum_{i\in \mathcal{S}}\delta_i w_i(1+\kappa_i)x_i R(x_i)-
\sum_{i\in\mathcal{S}}w_ix_i R(x_i)\nonumber\\
&=\sum_{i\in \mathcal{S}}\delta_i w_i \{1+\sum_{j\in \mathcal{S}}\frac{w_j x_j}{w_ix_i}(1-\delta_j)d_{ij} \} x_iR(x_i)-
\sum_{i\in\mathcal{S}}w_ix_i R(x_i)\nonumber\\
&=
\sum_{i\in \mathcal{S}}
\delta_i d_{ij}R(x_i) \sum_{j\in \mathcal{S}}w_j x_j(1-\delta_j)-
\sum_{i\in\mathcal{S}}(1-\delta_i)w_ix_i R(x_i)\nonumber\\
&=
\sum_{j\in \mathcal{S}}w_j x_j(1-\delta_j) \sum_{i\in \mathcal{S}}
\delta_i d_{ij}R(x_i) -
\sum_{i\in\mathcal{S}}(1-\delta_i)w_ix_i R(x_i)\nonumber\\
&=\textcolor{blue}{-} \sum_{i\in\mathcal{S}}
(1-\delta_i)w_i x_i
\{R(x_i)-R(x_{i(1)})\}.\label{asy:R}
\end{align}
Under  Assumption \ref{ass:regular}, \eqref{asy:R} can be bounded by $o_p(1)$  by the following argument:
\begin{align}
   | \sum_{i\in\mathcal{S}}
(1-\delta_i)w_i x_i
\{R(x_i)-R(x_{i(1)})\} |&\leq 
\sum_{i\in\mathcal{S}}
|(1-\delta_i)w_i x_i|\cdot | \{R(x_i)-R(x_{i(1)})\} |\nonumber \\
&\leq 
C_3\sum_{i\in\mathcal{S}}
|(1-\delta_i)w_i x_i| \cdot| x_i-x_{i(1)}|,\label{eq:diff}
\end{align}
where $| x_i- x_{i(1)}|$ is termed as the matching discrepancy, uniformly bounded by $O_p(n^{-1})$ as shown in Lemma \ref{lemma:discrep} (provided immediately below). 

Hence, under Assumptions \ref{assump:MAR}--\ref{ass:regular},  the term in \eqref{asy:R} is bounded by $O_p(n^{-1}N)$. Then, we have
\begin{equation}
   \sum_{i\in \mathcal{S}}\delta_i w_i(1+\kappa_i)x_i R(x_i)=
\sum_{i\in\mathcal{S}}w_ix_i R(x_i)+O_p(n^{-1}N). 
\label{eq:asym_equal}
\end{equation}
It follows from (\ref{eq:asym_equal}) that (\ref{4}) holds asymptotically:
\begin{align*}
    n^{1/2}N^{-1}\widehat{T}_{y,I}&=n^{1/2}N^{-1}\sum_{i\in S} w_i\delta_i(1+\kappa_i)y_i\\
    &=n^{1/2}N^{-1}\sum_{i\in S}w_{i}\left[ x_{i}{R}(x_i)+\delta_{i}(1+\kappa_{i})\{y_{i}-x_{i}{R}(x_i)\}\right]+o_p(1).
\end{align*}

\begin{lemma}\label{lemma:discrep}
Under Assumption  \ref{ass:regular} and let $|V|=n|X_i-X_{i(1)}|$, then the limiting distribution of $V$ is non-degenerate. Hence $|X_i-X_{i(1)}|$ is bounded by $O_p(n^{-1})$ uniformly.
\end{lemma}
To start, we consider the conditional probability that unit $j$ is the nearest neighbor of  $X_i=x_i$, i.e., $i(1)=j$
\begin{equation*}
    P(i(1)=j\mid X_j = x_j)=
    P(|X-x_i|\geq |x_j-x_i|)^{n-1}.
\end{equation*}
Since $x_i$ are i.i.d. $f_X$ ($i=1,\cdots,n$), the marginal probability that unit $j$ is a match for any nonrespondent is $1/n$. Therefore, the distribution of $X_j$ conditional on being the nearest neighbor ($1^{st}$ closest match) is
\begin{align*}
    f_{X_j\mid i(1)=j}&=
    \frac{P(i(1)=j\mid X_j=x_j)f_X(x_j)}{P(i(1)=j)}\\
    &=
    nf_X(x_j) P(|X-x_i|\geq |x_j-x_i|)^{n-1}.
\end{align*}
Next, we transform $X_j$ to a new random variable by $V=n( X_i - X_j)=n( X_i - X_{i(1)})$
\begin{align*}
    f_{V}(v)&=n f_X(x_i-\frac{v}{n})
    P(|X-x_i|\geq \frac{|v|}{n})^{n-1}\cdot \frac{\partial X_j}{\partial V}\\
    &=f_X(x_i-\frac{v}{n})
    \left\{
    1 - P(|X-x_i|\leq \frac{|v|}{n})
    \right\}^{n-1}.
\end{align*}
Then, we can show that
\begin{align*}
    \left\{
    1 - P(|X-x_i|\leq \frac{|v|}{n})
    \right\}^{n-1}&=
    \left\{
    1 - P(|X-x_i|\leq \frac{|v|}{n})
    \right\}^{n}\{1+o(1)\}\\
    &=\left\{
    1-\frac{A(x_i,v,n)}{n}
    \right\}^{n}\{1+o(1)\},
\end{align*}
where $A(x_i,v,n)$ can be specified as
$$
A(x_i,v,n)=nP(|X-x_i|\leq {|v|}/{n}).
$$
Note that 
$$
P(|X-x_i|\leq {|v|}/{n})=
\int_0^{{|v|}/{n}}
\int_{\mathbb{S}} f_X(x_i+rw)
\lambda_{\mathbb{S}}(dw)dr,
$$
where $\mathbb{S}=\{\omega\in \mathbb{R}:\|\omega\|=1\}$ is the unit $k$ sphere with $\lambda_{\mathbb{S}}$ as its surface measure. Therefore, by l'Hopital's rule, we can obtain the limit of $A(x_i,v,n)$ as $n\rightarrow \infty$,
\begin{align*}
\lim_{n\rightarrow \infty}
\frac{P(|X-x_i|\leq {|v|}/{n})}{1/n}&=
\lim_{n\rightarrow \infty}
\frac{-|v|/n^2 \int_{\mathbb{S}} f_X(x_i+|v|\omega/n)
\lambda_{\mathbb{S}}(dw)}{-1/n^2}\\
&=|v|\int_{\mathbb{S}} f_X(x_i)
\lambda_{\mathbb{S}}(dw).
\end{align*}
Hence, we have the limiting distribution of $V$ as 
\begin{equation}
    f_V(v)=
    f_X(x_i)\exp\left\{-|v|\int_{\mathbb{S}} f_X(x_i)\lambda_{\mathbb{S}}(dw)\right\}.
\end{equation}
In addition, we know that $f_X$ is bounded in the support $\mathbb{X}$ by Assumption \ref{ass:regular} and $|V|=n|X_i-X_{i(1)}|$ is also non-degenerate by simple algebra, i.e., $|V|=O_p(1)$, $|x_i-x_{i(1)}|=O_p(n^{-1})$ uniformly.

\section{Proof of Theorem \ref{theorem:normal}}\label{subsec:ProofT1}
By Theorem 1 of \citet{yang2020asymptotic}, we have 
\begin{equation}
n^{1/2}N^{-1}(\widehat{T}_{y,I}-T_{y})=D_{N}^*+B_{N},
\end{equation}
where 
\[
D_{N}^{*}=n^{1/2}N^{-1}\left[\sum_{i\in S}w_{i}\{m_{i}+\delta_{i}(1+\kappa_{i})e_{i}\}-T_{y}\right],
\]
\[
B_{N}=n^{1/2}N^{-1}
\left[ \sum_{i\in\mathcal{S}}
(1-\delta_i)w_i x_i
\{R(x_{i(1)}) - R(x_i)\}\right],
\]
 $m_{i}=x_{i}R(x_i)$, and $e_{i}=y_{i}-m_{i}$. Under Assumption \ref{ass:regular} and Lemma \ref{lemma:discrep}, $B_{N}$ is uniformly bounded by $O_p(n^{-1})$ for any $x_i,x_{i(1)}\in\mathbb{X}$. Thus, $B_N$ is asymptotically negligible. Since $T_{y}=\sum_{i=1}^{N}(m_{i}+e_{i})$, we can express the first term as
\[
D_{N}^{*}=n^{1/2}N^{-1}\sum_{i=1}^{N}\left(I_{i}w_{i}-1\right)m_{i}+n^{1/2}N^{-1}\sum_{i=1}^{N}\left\{ I_{i}w_{i}\delta_{i}(1+\kappa_{i})-1\right\} e_{i},
\]
where $I_i=1$ if $i\in\mathcal{S}$. The expression of $D_N^*$ holds true even when $f=n/N$ is non-negligible. 
Next, we verify that the covariance of the two terms in $D_{N}^{*}$ equals zero.
\begin{align*}
   &\text{cov}\left\{\sum_{i=1}^{N}\left(I_{i}w_{i}-1\right)m_{i},\sum_{i=1}^{N}\left\{ I_{i}w_{i}\delta_{i}(1+\kappa_{i})-1\right\} e_{i}\right\}\\
   &=E\left\{\sum_{i=1}^{N}\left(I_{i}w_{i}-1\right)m_{i}\times \sum_{i=1}^{N}\left\{ I_{i}w_{i}\delta_{i}(1+\kappa_{i})-1\right\} e_{i}\right\}\\
   &-
   E\left\{\sum_{i=1}^{N}\left(I_{i}w_{i}-1\right)m_{i}\right\}
   E\left\{\sum_{i=1}^{N}\left\{ I_{i}w_{i}\delta_{i}(1+\kappa_{i})-1\right\} e_{i}\right\}\\
  &= E\left[\sum_{i=1}^{N}\left(I_{i}w_{i}-1\right)m_{i}\times E_\zeta\left\{ \sum_{i=1}^{N}\left\{ I_{i}w_{i}\delta_{i}(1+\kappa_{i})-1\right\} e_{i}\mid x_i\right\}\right]\\
   &=0,
\end{align*}
where $E_\zeta(e_i\mid x_i)=0$ by definition. Therefore, the asymptotic variance is essentially equal to 
\[
\text{var}(D_{N}^*)=V^{m}+V^{e},
\]
where
\begin{align*}
    V^m&=
    \frac{n}{N^2}\text{var}\left\{ 
    \sum_{i=1}^N(I_iw_i-1)m_i
    \right\} \\
    &=
    \frac{n}{N^2}\text{var}_{}\left\{ E_p
    \sum_{i=1}^N(I_iw_i-1)m_i
    \right\}+
    \frac{n}{N^2}{E}_{}\left\{ \text{var}_p
    \sum_{i=1}^N(I_iw_i-1)m_i
    \right\}\\
    &= \frac{n}{N^2}{E}_{}\left\{ \text{var}_p
    \left(\sum_{i\in S}w_{i}m_{i}-\sum_{i=1}^{N}m_{i}\right)
    \right\},
\end{align*} and
\begin{align*}
    V^e&= \frac{n}{N^2}\text{var}\left[\sum_{i=1}^{N}\left\{ I_{i}w_{i}\delta_{i}(1+\kappa_{i})-1\right\} e_{i}\mid x_{i}\right]\\
    &=\frac{n}{N^2}\text{var}\left[E\sum_{i=1}^{N}\left\{ I_{i}w_{i}\delta_{i}(1+\kappa_{i})-1\right\} e_{i}\mid x_{i}\right]+
    \frac{n}{N^2}E_{}\left[\text{var}\sum_{i=1}^{N}\left\{ I_{i}w_{i}\delta_{i}(1+\kappa_{i})-1\right\} e_{i}\mid x_{i}\right]\\
    &=
    \frac{n}{N^2}E_{}\left[\sum_{i=1}^{N}\left\{ I_{i}w_{i}\delta_{i}(1+\kappa_{i})-1\right\}^2 E(e_i^2\mid x_i)\right]\\
    &=\frac{n}{N^2}
    \sum_{i=1}^N\{
    w_i\delta_i(1+\kappa_i)^2+
    1-
    2\delta_i(1+\kappa_i)\}E(e_i^2\mid x_i),
\end{align*}
where we can show $V^e=O(1)$
\begin{eqnarray*}
    \frac{n}{N^2}
    \sum_{i=1}^N\{
    w_i\delta_i(1+\kappa_i)^2+
    1-
    2\delta_i(1+\kappa_i)\}E(e_i^2\mid x_i)
    &=&
    \frac{n}{N^2}
    \sum_{i=1}^N  w_i\delta_i(1+\kappa_i)^2{\sigma}_e^2(x_i)\\
    &&+
    \frac{n}{N^2}
    \sum_{i=1}^N 
    \{ 1-
    2\delta_i(1+\kappa_i)\}{\sigma}^2_e(x_i),
\end{eqnarray*}
with $\sigma_e^2(x_i)=E(e_i^2\mid x_i)$. The first term can be characterized under Assumptions \ref{assump:design} and \ref{ass:regular}
$$
\frac{n}{N^2}
    \sum_{i=1}^N  w_i\delta_i(1+\kappa_i)^2{\sigma}_e^2(x_i)
    = \frac{n}{N^2}\cdot \frac{CN}{n}\sum_{i=1}^N \delta_i (1+\kappa_i)^2\sigma_e^2(x_i)=O(1),
$$
where $\kappa_i =\sum_{j\in\mathcal{S}}w_jx_j/(w_ix_i)(1-\delta_j)d_{ij}$ is bounded by
$$
\frac{C_1\min_{x\in \mathbb{X}}|x|}{C_2\max_{x\in \mathbb{X}}|x|}\Tilde{\kappa}_i
\leq 
\kappa_i\leq \frac{C_2\max_{x\in \mathbb{X}}|x|}{C_1\min_{x\in \mathbb{X}}|x|}\Tilde{\kappa}_i,
$$
where $\Tilde{\kappa}_i=\sum_{j\in\mathcal{S}}(1-\delta_j)d_{ij}$ is bounded by $O(1)$ \citep[p.17]{abadie2006large, yang2020asymptotic}. Next, the second term can be bounded similarly
\begin{equation*}
     \frac{n}{N^2}
    \sum_{i=1}^N 
    \{ 1-
    2\delta_i(1+\kappa_i)\}{\sigma}_e^2=
    \frac{n}{N^2}
    \sum_{i=1}^N\{1+O_p(1)\}\sigma_e^2=
    O(nN^{-1}).
\end{equation*}
Note that when $n/N$ is negligible, the term $V_e$ is then dominated by the first term ${n}{N^{-2}}\sum_{i=1}^N w_i\delta_i(1+\kappa_i)^2E(e_i^2\mid x_i)$.

\end{document}